\documentclass{aa}

\usepackage{graphicx}
\usepackage{txfonts}
\usepackage{amsmath}
\usepackage{url}
\usepackage[colorlinks,linkcolor=blue,citecolor=blue,urlcolor=blue]{hyperref}

\newcommand{\mbh}{\ensuremath{M_{\rm BH}}}
\newcommand{\m}{\ensuremath{M_{\star}}}
\newcommand{\msun}{\ensuremath{M_\odot}}
\newcommand{\logm}{\ensuremath{\log\,M_{\star}/M_\odot}}
\newcommand{\logmbh}{\ensuremath{\log\,M_{\rm BH}/M_\odot}}
\newcommand{\lbol}{\ensuremath{L_{\rm bol}}}
\newcommand{\loglbol}{\ensuremath{\log\,L_{\rm bol}/{\rm erg\,s^{-1}}}}
\newcommand{\edd}{\ensuremath{\lambda_{\rm Edd}}}
\newcommand{\rel}{\ensuremath{M_{\rm BH}\text{--}M_{\star}}}
\newcommand{\ergs}{\ensuremath{{\rm erg\,s^{-1}}}}
\newcommand{\kms}{\ensuremath{{\rm km\,s^{-1}}}}
\newcommand{\ha}{H$\alpha$}

\newcommand{\oiii}{[O\,{\sc iii}]}
\newcommand{\nii}{[N\,{\sc ii}]}
\newcommand{\sii}{[S\,{\sc ii}]}
\newcommand{\cigale}{\texttt{CIGALE}}

\begin{document}

   \title{A prevalent population of normal-mass central black holes in high-redshift massive galaxies}

   \author{
   Junyao Li\inst{1,2,3}\corrauth{lijunyaox@gmail.com}
   \and 
   Yue Shen\inst{3,4}
   \and 
   Ming-Yang Zhuang\inst{3}
   }

   \institute{
   Finnish Centre for Astronomy with ESO (FINCA), University of Turku, 20014 Turku, Finland
   \and 
   Department of Physics, University of Helsinki, Gustaf Hällström katu 2, 00560 Helsinki, Finland
   \and
   Department of Astronomy, University of Illinois at Urbana-Champaign, Urbana, IL 61801, USA
   \and 
   National Center for Supercomputing Applications, University of Illinois Urbana-Champaign, Urbana, IL 61801, USA}

   \date{Received ...; accepted ...}

  \abstract
   {Understanding the co-evolution of supermassive black holes (SMBHs) and their host galaxies is crucial for explaining how feedback from accreting black holes regulates star formation and shapes the growth of galaxies across cosmic time. However, measurements of this co-evolution in the early Universe, as traced by the black hole mass--stellar mass relation (\mbh-\m), are subject to strong selection biases: SMBHs that are overly luminous and massive relative to their hosts are preferentially detected, particularly in the more abundant low-mass galaxies. To mitigate these biases, we conduct a systematic search for broad-line active galactic nuclei (AGN) in 117 spectroscopically confirmed massive galaxies at $z\sim3-5$, using deep JWST imaging and spectroscopic surveys conducted to date. We identify 20 moderate-luminosity AGNs with a mean \mbh/\m\ of $\sim0.1\%$, consistent with the local value. By reconstructing their mass ratios back to $z\sim6$, we infer that a substantial population of SMBHs with normal mass ratios was already in place within the first billion years of cosmic history, but has been largely overlooked by existing surveys.  Together with the overmassive black holes reported in previous studies, our findings suggest diverse pathways for SMBH formation and growth in the early Universe. In addition, we find comparable \mbh/\m\ ratios in star-forming and quiescent host galaxies, suggesting that the formation of massive quiescent galaxies does not necessarily require an overmassive black hole, contrary to common theoretical expectations.}

   \maketitle
   \nolinenumbers

\section{Introduction}
\label{sec:intro}

The masses of supermassive black holes (SMBHs) in local galaxies correlate tightly with stellar velocity dispersion and bulge mass \citep{Kormendy2013, Greene2020}. These correlations are widely interpreted as evidence that SMBHs and their host galaxies grow together. However, the origins of SMBH seeds and the processes that established the local relations remain unresolved \citep{Volonteri2010}. In theoretical models, outflows and jets from active galactic nuclei (AGNs) inject energy and momentum into the interstellar medium. This feedback can regulate both black hole accretion and star formation, coupling SMBH growth to galaxy evolution \citep[e.g.][]{Fabian2012, King2015}. Alternatively, the correlations could arise through the statistical averaging of repeated mergers, without a causal link between the two growth processes \citep{Peng2007, Jahnke2011}. Cosmological simulations and semi-analytic models predict different evolutionary tracks for the black hole mass–stellar mass (\rel) relation. These differences reflect their assumptions about black hole seeding, accretion, merger and feedback. The predictions diverge most strongly at $z\gtrsim3$, where models disagree on whether SMBHs are overmassive or undermassive relative to the local \rel\ relation \citep{Habouzit2021, Habouzit2022b, Ding2022, Dattathri2024, Bhowmick2024}. Measuring the mass relation at high redshift therefore provides a direct test of SMBH formation and coevolution mechanisms with their host galaxies.

The \textit{James Webb Space Telescope} (JWST) has enabled the detection of broad-line AGNs (BLAGNs) and their host galaxies at $z>4$, with many faint systems comparable to local AGNs \citep{Harikane2023, Kocevski2023, Kokorev2023, Maiolino2023, Ubler2023, Matthee2024, Greene2024, Taylor2024, Zhuang2026}. JWST has also provided the first detections of stellar emission from the hosts of luminous quasars at $z\sim6$ \citep{Ding2023, Stone2024, Yue2024}. These samples span more than three orders of magnitude in AGN luminosity and host stellar mass. Their SMBHs lie, on average, 1--2 dex above the local relation, with $\mbh/\m\gtrsim1-10\%$ \citep{Pacucci2023, Maiolino2023, Juodzbalis2026census}. These apparently overmassive SMBHs have been interpreted as evidence that episodes of super-Eddington accretion or heavy black hole seeds played a dominant role in the early growth of SMBHs \citep[e.g.][]{Natarajan2024, Schneider2023, Trinca2026, Jeon2025}. 

However, selection bias complicates the interpretation of these results. Flux-limited surveys preferentially detect luminous AGNs and favour overmassive SMBHs at a given host stellar mass \citep{Li2021mass, Li2022}. This statistical effect, known as the Lauer bias \citep{Lauer2007}, could largely explain the elevated and flattened \rel\ relation inferred at early epochs \citep{Schulze2014, Volonteri2023, Li2024}. Shallow, wide-area quasar surveys primarily detect the most luminous and massive SMBHs in massive galaxies \citep{Fan2023}. Deep JWST pencil-beam surveys reach much fainter AGNs, but can still preferentially detect overmassive SMBHs in the far more numerous low-mass galaxies \citep{Li2024}. Massive hosts containing SMBHs of similar mass and luminosity, but lower $\mbh/\m$, are much rarer within these small survey volumes. Forward modelling of JWST selection effects indicates that the observed offsets may remain consistent with an intrinsic relation closer to the local one but with larger scatter \citep{Li2024, Silverman2025, Ziparo2026, Koudmani2026}. Stacking analyses also suggest that lower-mass SMBHs closer to the local relation remain undetected in individual spectra \citep{Geris2026, Brooks2026}. 

The high stellar mass regime offers an opportunity to constrain the intrinsic mass relation at early cosmic time where SMBHs consistent with the local relation are detectable in JWST surveys \citep{Li2024}. We refer to these objects as ``normal-mass'' SMBHs throughout this paper. Their counterparts in low-mass galaxies have lower black hole masses and can fall below the detection limit assuming similar Eddington ratio distributions. Targeting massive galaxies therefore extends the search to a regime in which detectable SMBHs need not be overmassive \citep{Li2024}. 

Constraining SMBH masses in massive high-redshift galaxies is also crucial for understanding galaxy quenching. JWST spectroscopy has confirmed a surprisingly abundant population of massive quiescent galaxies at $z>3$. These galaxies formed most of their stars within the first billion years of cosmic history and subsequently underwent rapid quenching \citep[e.g.][]{Carnall2023nat, Carnall2024, Glazebrook2024, Nanayakkara2025, deGraaff2024qs}. Cosmological simulations struggle to reproduce their observed number densities \citep{Valentino2023, Baker2024, Weller2024}. In simulations that successfully produce such galaxies, quenching at $z\gtrsim3$ is driven by AGN feedback. Their SMBHs are seeded early, grow rapidly and become overmassive relative to their host galaxies \citep{Weller2024, Dattathri2024}. In the local Universe, quiescence correlates more strongly with black hole mass, central stellar velocity dispersion than with total stellar mass. Quiescent galaxies also host more massive SMBHs at fixed $\m$ \citep{Terrazas2016, Terrazas2017, MartinNavarro2018, Bluck2020, Bluck2022, Piotrowska2022}. These trends have been attributed to the integrated energy output of SMBHs \citep{Terrazas2020}. However, although JWST observations have revealed widespread AGN activity, outflows and evidence for AGN-driven gas removal in the early Universe \citep{Ito2022, DEugenio2023, Belli2024, Park2024, Bugiani2025, Valentino2025}, whether the earliest quiescent galaxies host overmassive SMBHs, as predicted by these simulations, has not been tested directly. 

Here, we search for BLAGNs in spectroscopically confirmed massive galaxies at $3\le z\le5$ using all publicly available NIRSpec spectroscopy from the DJA archive \citep{msaexp, deGraaff2024}. The  large combined survey area ($\sim 0.2\ \deg^2$) is crucial for identifying rare massive galaxies hosting normal-mass SMBHs, while the detection of broad emission lines enables single-epoch virial black hole mass estimates \citep{Vestergaard2006, Shen2024RM}. We first test whether high-redshift SMBHs are universally overmassive by measuring $\mbh/\m$ in the regime where normal-mass SMBHs are predicted to be detectable. We then investigate whether the earliest massive quiescent galaxies host the most massive SMBHs, as predicted by simulations in which AGN feedback drives quenching. 

Section~\ref{sec:methods} describes the data, sample selection, spectral energy distribution (SED) fitting, spectral decomposition, and black hole mass estimation methods. Section~\ref{sec:results} presents the properties of BLAGNs and their host galaxies, including their distribution in the \rel\ plane. Section~\ref{sec:discussion} discusses the implications of our results for black hole seeding and galaxy quenching. We summarize our conclusions in Section~\ref{sec:conclusions}. Throughout this paper, we adopt a flat $\Lambda$CDM cosmology with $H_0=70$ \kms\ Mpc$^{-1}$ and $\Omega_{\rm m}=0.3$.

\section{Data and methods}
\label{sec:methods}

To identify BLAGNs in massive galaxies at high redshift, we first assembled a sample of spectroscopically confirmed galaxies at $3<z<5$ with significant \ha\ detections. We retrieved publicly available NIRSpec multi-shutter array (MSA) spectra and corrected their multiband photometry for emission-line contamination. We then estimated stellar masses using SED models that account for both stellar and AGN emission. Finally, we identified BLAGNs among the massive galaxies based on the presence of broad \ha\ emission. The following subsections describe the sample selection, data processing, and analysis procedures. 

\subsection{NIRSpec spectroscopy and parent sample}
\label{sec:spec}

We obtained spectra from v4.4 of the DAWN NIRSpec Archive (DJA\footnote{\url{https://dawn-cph.github.io/dja/}}), which provides uniformly reduced NIRSpec MSA observations using \texttt{msaexp} \citep{msaexp, Heintz2025, deGraaff2024}. In addition to the standard JWST calibration steps, \texttt{msaexp} corrects for MSA bar shadows and subtracts the background by differencing nodded exposures. It fits the cross-dispersion profile using a wavelength-dependent trace and optimally extracts the one-dimensional spectrum \citep{Horne1986}. The pipeline also applies a path-loss correction based on the source position and size within the shutter. Spectroscopic redshifts are determined through template fitting, with quality grades assigned by visual inspection; grade~3 indicates a robust redshift. The v4.4 release also provides a emission-line catalog measured from the reduced spectra.

We selected sources with robust grade-3 spectroscopic redshifts at $3<z<5$ and at least one medium- or high-resolution NIRSpec grating spectrum (G140M/H, G235M/H or G395M/H) covering \ha\ with $\mathrm{S/N}>5$. The lower redshift limit defines the high-redshift population targeted in this study, while the upper limit ensures adequate rest-frame optical coverage with NIRCam photometry, which is important for reliable stellar mass estimates. These criteria yielded 2421 sources.

Some sources with grade-3 redshifts in DJA v3 are either absent from v4.4 or lack redshift measurements in that release. We therefore applied the same selection criteria to the v3 catalog and identified 51 additional sources. Visual inspection confirmed the reliability of their redshifts and revealed no obvious issues with the reduced spectra. Our parent galaxy sample thus comprises 2472 galaxies.

The parent sample includes observations from 27 NIRSpec programs. The largest contributions come from RUBIES \citep[GO 4233;][]{deGraaff2024}, JADES \citep[GTO 1180, 1181, 1286;][]{Eisenstein2023, DEugenio2025dr3}, and EXCELS \citep[GO 3543;][]{Carnall2024}. Other major contributors include GO 3567 (PI: F. Valentino) and the NIRSpec Wide GTO survey \citep[GTO 1211--1215;][]{Maseda2024}. These programs have different scientific objectives and adopt different target selection strategies for shutter allocation. Consequently, the parent sample is heterogeneous, and its selection function cannot be readily quantified. 

\subsection{Photometric catalogs}
\label{sec:phot}

We estimated stellar masses using publicly available multiband photometric catalogs covering six deep extragalactic fields. For GOODS-S and GOODS-N, we used JADES DR5 \citep{Robertson2026}, while for EGS, we used CEERS DR1 \citep{Cox2026}. For the PRIMER regions in COSMOS and UDS, we used the ULTIMATE-deblending catalogs \citep{SunH2026}. In COSMOS, we additionally used the COSMOS-Web catalog for sources outside the PRIMER footprint \citep{Shuntov2025}. For Abell~2744, we used UNCOVER DR3 \citep{Suess2024}. All catalogs use deep NIRCam imaging for source detection and provide photometry in the available NIRCam bands. Depending on the field, they also include photometry from JWST/MIRI, HST, ground-based facilities, and \textit{Spitzer}.

To estimate stellar masses, we adopted total fluxes from each photometric catalog. For JADES DR5, we used Kron fluxes measured from images convolved to the F444W point-spread function and corrected to total fluxes. For CEERS DR1, we used aperture-corrected Kron fluxes. For PRIMER, we used the ``best'' total fluxes in the \textit{HST} and \textit{JWST} bands and TPHOT fluxes in the ground-based and \textit{Spitzer}/IRAC bands. For COSMOS2025, we used SE++ model fluxes in all bands, while for UNCOVER, we adopted the DR3 SUPER color-aperture fluxes corrected to total fluxes. We cross-matched the spectroscopic sample with these catalogs using a matching radius of 0.3\arcsec. To ensure reliable SED decomposition and stellar mass estimates, we required at least one photometric measurement at rest-frame wavelengths $\le3000$~\AA\ and another at $\ge6000$~\AA. After applying these criteria, 2164 objects remained, with a median of 20 photometric bands per object used for SED fitting.

\subsection{Spectrophotometric calibration and emission-line correction of the photometry}
\label{sec:calib}

Strong emission lines can substantially contaminate the broadband photometry of AGN host galaxies, particularly broad emission lines that are not self-consistently modeled by current SED-fitting codes. We therefore used the available PRISM and grating spectra to measure and subtract the emission-line contributions before fitting the photometry. This procedure requires the spectra and photometry to be on the same flux scale. We thus first calibrated the spectra against the photometry to correct for slit losses, which can be substantial when sources are offset from their shutter centers. Although \texttt{msaexp} applies a path-loss correction, its assumed source profile may not fully account for additional wavelength-dependent slit losses in extended galaxies.

PRISM spectra were available for 1676 of the 2164 parent sources. When multiple PRISM spectra were available for a given source, we selected the one with the best wavelength coverage and continuum S/N. We computed synthetic photometry for bands in which the spectrum covered at least 90\% of the filter response. For bands with photometric $\mathrm{S/N}\ge3$, we fitted the ratio of observed to synthetic fluxes with a polynomial as a function of wavelength. We selected the polynomial degree (0--5) by minimizing the Bayesian information criterion (BIC) value. We visually inspected all fits and manually refitted 18 sources by either fixing the polynomial degree or excluding problematic bands. The median correction factors at the pivot wavelengths of F115W, F150W, F277W, and F444W were 1.12, 1.12, 1.14, and 1.21, respectively.

We calibrated each medium and high-resolution spectrum using the same procedure. When a PRISM spectrum was available, we computed synthetic photometry in multiple bands from the calibrated PRISM spectrum and used these measurements to calibrate the grating spectra. Otherwise, we calibrated the grating spectra directly against the observed photometry, which typically provided fewer bands for calibration. In the latter case, we restricted the polynomial degree to $\le2$ to avoid overfitting.

We then fitted the emission lines in the calibrated spectra using Gaussian profiles and local linear continua. The fits included major line complexes spanning Ly$\alpha$ to Pf$\beta$. Lines within each complex shared the same velocity and width, with fixed doublet ratios imposed where appropriate. We allowed additional broad components for the Balmer, Paschen, Mg\,\textsc{ii}, C\,\textsc{iv}, and Ly$\alpha$ lines, as well as outflow wings for \oiii. Broad components were retained only if they improved the fit by $\Delta\mathrm{BIC}>10$. For lines detected at $\mathrm{S/N}>3$ in the grating spectra, we adopted the measurement with the highest S/N. For the remaining weakly detected lines, we used measurements from the PRISM spectra where available.

We integrated the best-fit emission line profiles through the filter transmission curve to estimate their contributions to each photometric band, which were then subtracted from the observed fluxes. We propagated uncertainties in both line fluxes and spectral flux calibration into photometric errors. We corrected the photometry for strong line contamination in 2041 sources. For the remaining 123 sources, reliable line subtraction was not possible, mainly because of insufficient spectral coverage. We therefore performed SED fitting using the original photometry with models that include nebular emission lines and AGN narrow emission lines with fixed line ratios, but do not account for AGN-powered broad emission lines (see Section \ref{sec:sed}).

\begin{table*}
\caption{Free parameters and priors of the \texttt{Prospector} fits.}
\label{tab:sedparams}
\centering
\begin{tabular}{ll}
\hline\hline
Parameter & Prior \\
\hline
$\log\,M_{\rm formed}/M_\odot$ & Uniform $[7, 12]$ \\
$r_1$, $r_2$ (0--10 and 10--100\,Myr bins) & Uniform $[-4, 2.5]$ \\
$r_3$--$r_6$ ($>$100\,Myr) & Student's $t$ ($\mu=0$, $\sigma=0.3$, $\nu=2$) \\
$\log\,Z/Z_\odot$ & Uniform $[-0.70, 0]$ \\
$\tau_V$ (all stars) & Uniform $[0, 2.5]$ \\
$\tau_{V,\rm birth}/\tau_V$ (stars $<10$\,Myr only) & Fixed at 1 \\
$f_{\rm AGN}$ (disk fraction of $L_\nu$ at 5100\,\AA) & Uniform $[0.0, 1.0]$ \\
$E(B-V)_{\rm polar}$ (SMC) & Uniform $[0.0, 0.5]$ \\
\hline
\end{tabular}
\tablefoot{SKIRTOR torus geometry fixed ($\tau_{9.7}=7$, $p=1$, $q=1$, $\Delta=40^\circ$, $R_{\rm out}/R_{\rm in}=20$, $M_{\rm cl}=0.97$, $i=30^\circ$ as adopted in \cigale); polar dust fixed at 100\,K with $\beta=1.6$. Galactic dust emission (\citealt{Draine2007}, fixed to \texttt{Prospector} default) is included only for sources with F1800W photometry.}
\end{table*}

\subsection{SED decomposition and stellar masses}
\label{sec:sed}

\begin{figure*}
\centering
\includegraphics[width=\hsize]{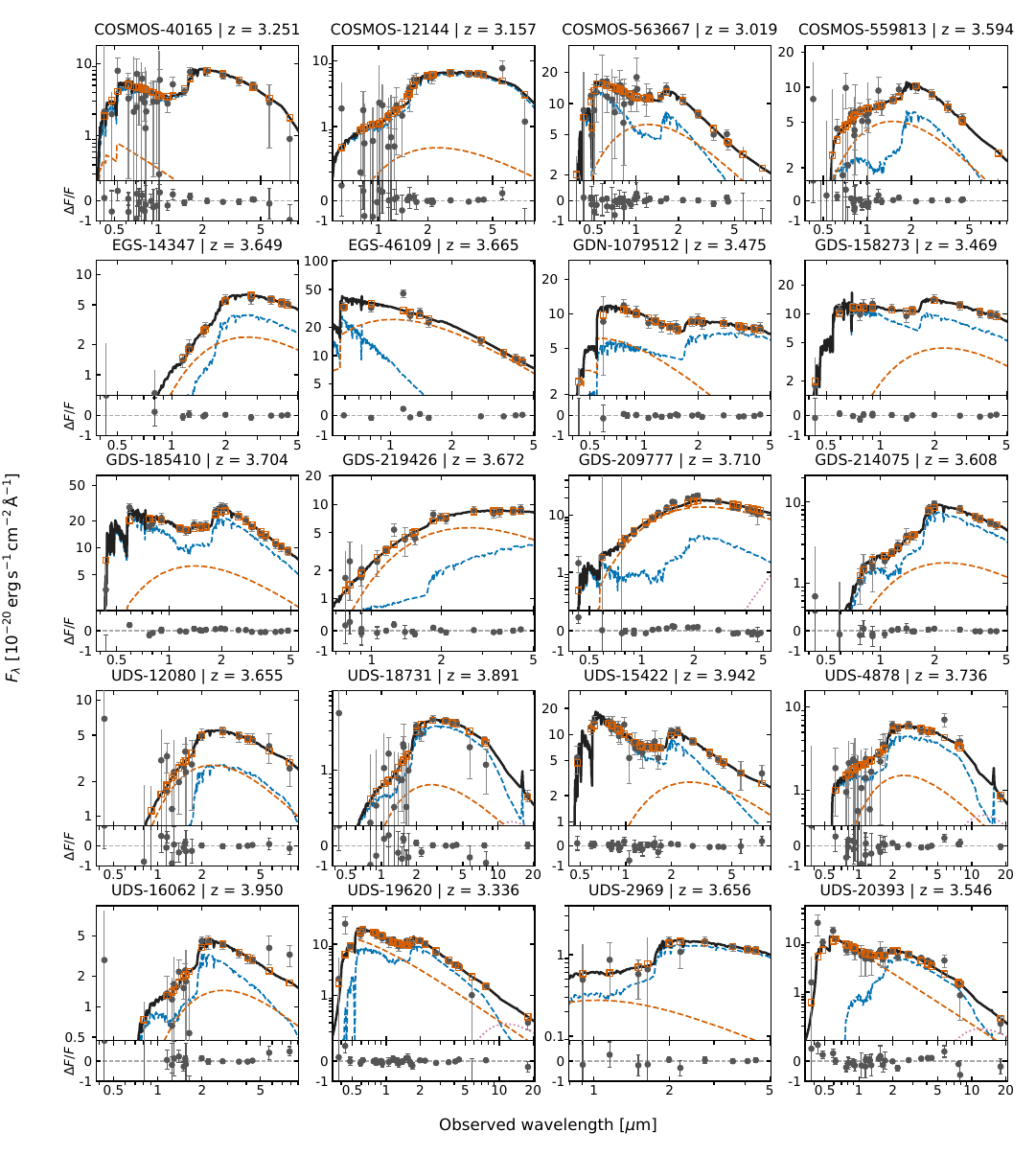}
\caption{Spectral energy distributions of the 20 BLAGNs in the final sample, decomposed using \texttt{Prospector} with the SKIRTOR AGN templates. Each panel shows the emission-line-corrected photometry with uncertainties, along with the best-fit total model (solid black line) and its stellar (dashed blue), AGN (dashed orange), and galactic dust (dotted pink, only for sources covered by F1800W) components. The lower panel shows the fractional fitting residuals.}
\label{fig:sed}
\end{figure*}

We fitted the photometry using \texttt{Prospector} \citep{Johnson2021}, which combines the Flexible Stellar Population Synthesis code \citep[FSPS;][]{Conroy2009, Conroy2010} with Bayesian inference to measure the stellar mass and star-formation history (SFH) of our sample. The built-in AGN module of \texttt{Prospector} is the CLUMPY torus model \citep{Nenkova2008a, Nenkova2008b}, which has no accretion-disk component. We therefore replaced it with the SKIRTOR AGN model \citep{Stalevski2012, Stalevski2016} as implemented in \cigale\ \citep{cigale, Yang2022}.  

The SKIRTOR model comprises three components: an accretion disk, a two-phase clumpy torus, and polar dust. The accretion-disk spectrum follows the modified broken power-law prescription of \citet{Schartmann2005}. To reduce parameter degeneracies, we fixed the torus optical depth and geometry to their default values in \cigale. We fixed the viewing inclination at $30^\circ$, corresponding to a type-1 line of sight appropriate for BLAGNs. Polar dust attenuates the accretion-disk emission following the Small Magellanic Cloud (SMC) extinction curve \citep{Prevot1984} and re-emits the absorbed energy as a modified blackbody with a temperature of $T=100$\,K and an emissivity index of $\beta=1.6$. The AGN normalization is controlled by the AGN fraction parameter, $f_{\rm AGN}$, defined as the fraction of the total model luminosity at rest-frame 5100\,\AA\ contributed by the AGN accretion disk. We adopted uniform priors on $f_{\rm AGN}$ over $[0.0, 1.0]$ and on the polar-dust $E(B-V)$ over $[0.0, 0.5]$.

For the stellar component, we adopted the MIST isochrones \citep{Choi2016, Dotter2016}, the C3K stellar spectral library, and a \citet{Chabrier2003} initial mass function (IMF). We adopted a non-parametric SFH with seven bins in lookback time: 0--10\,Myr, 10--100\,Myr, four logarithmically spaced bins between 100\,Myr and $0.9\,t_{\rm max}$, and a final bin from $0.9\,t_{\rm max}$ to $t_{\rm max}$, where $t_{\rm max}$ is the cosmic time elapsed between $z=20$ and the observed redshift. The SFH is described by the total stellar mass formed and the six logarithmic star formation rate (SFR) ratios between adjacent bins $r_i=\log\,\mathrm{SFR}_i/\mathrm{SFR}_{i+1}$.

The choice of SFH prior is particularly important in non-parametric SED fitting, as rest-frame UV/optical photometry provides only weak constraints on the detailed SFH, especially at early times. We therefore adopted a hybrid prior that treats recent and earlier star formation differently. For $r_3$--$r_6$, which describe the SFH more than 100\,Myr ago, we adopted the continuity prior of \citet{Leja2019}, parameterized as a Student's $t$-distribution centered at zero with a scale of 0.3 and $\nu=2$. This prior is motivated by the expectation that galaxy SFRs generally vary smoothly over the longer timescales probed by the older bins. It regularizes the poorly constrained early SFH by suppressing large, unconstrained fluctuations between adjacent bins, which could otherwise lead to spuriously accumulated old stellar masses. Nevertheless, its heavier tails compared to a Gaussian distribution allow abrupt changes in SFR, including rapid early star formation followed by quenching, when supported by the data. Non-parametric SFHs with this continuity prior have been shown to recover stellar masses and SFRs of simulated galaxies more accurately than commonly used parametric SFHs \citep{Carnall2019, Leja2019, Lower2020}. 

For $r_1$ and $r_2$, which control the SFRs over the past 0--10\,Myr and 10--100\,Myr, we instead adopted uniform priors over $[-4, 2.5]$. This choice allows greater flexibility in the recent SFH, which is crucial for determining the current star-forming or quiescent state of BLAGN hosts. Applying the continuity prior to these bins could bias against rapid quenching by favoring similar SFRs in adjacent bins, whereas identifying such rapid transitions is a central goal of this study.

We adopted uniform priors on stellar metallicity and stellar mass over $-0.70 \leq \log\,Z/Z_\odot \leq 0$ and $7.0 \leq \logm \leq 12.0$, respectively. We modeled dust attenuation using the \citet{Calzetti2000} attenuation curve, with a uniform prior on the $V$-band optical depth over $0 \leq \tau_V \leq 2.5$. Stars younger than 10\,Myr were subject to additional birth-cloud attenuation \citep{CharlotFall2000}, modeled with a power-law curve, $\tau \propto \lambda^{-1}$. The $V$-band optical depth of this additional component was fixed to $\tau_V$, such that young stars experience twice the total $V$-band attenuation of older stars. We neglected galactic dust emission except for sources with MIRI F1800W coverage, the only band with rest-frame wavelength exceeding $\sim2\,\mu$m in our adopted photometric catalog, as its contribution is expected to be negligible over rest-frame UV/optical wavelengths.

For sources with reliable emission-line subtraction (Section \ref{sec:calib}), we performed SED fitting without including emission-line components. For the remaining sources, we included both stellar nebular emission and AGN narrow-line emission in \texttt{Prospector}. However, the current version of \texttt{Prospector} does not support self-consistent modeling of AGN broad emission lines. The AGN narrow-line model also adopts fixed empirical line ratios from \cite{Richardson2014}, which may not fully capture the diversity of narrow-line emission in our sample. Nevertheless, we verified that including or excluding the AGN narrow-line component has a negligible impact on the inferred stellar masses.

The adopted parameter ranges are listed in Table~\ref{tab:sedparams}.
In the fitting, We added a 10\% fractional flux uncertainty in quadrature to the catalog photometric uncertainties to account for potential systematic uncertainties not captured by the reported errors, including color gradients, aperture corrections, imperfect spectrophotometric calibration, and residual emission-line contamination. We sampled the posterior distributions using the nested sampling code \texttt{dynesty} \citep{Speagle2020} with 1000 live points. We adopted the posterior median and 16th--84th percentiles as our fiducial estimates and uncertainties of the stellar mass and SFRs (averaged over the last 100\,Myr). Our SED fitting yielded 123 galaxies with $\logm\gtrsim10$ and reduced $\chi^2<5$, constituting the massive galaxy sample used for subsequent BLAGN search. Figure~\ref{fig:sed} illustrates our SED decomposition procedure using the final BLAGN sample described in Section~\ref{sec:specfit}.

\subsection{Constructing continuum models used for spectral fitting}
\label{sec:continuum}

Massive galaxies at $z\sim3-5$ often exhibit strong Balmer absorption features, which can affect the observed \ha\ profile and partially mask its broad wings. We therefore adopted the combined stellar and AGN continuum from the SED fits as the continuum template for identifying broad emission lines in the grating spectra. This approach accounts for stellar absorption features that cannot be adequately modeled by a simple power-law continuum.

The SED fits used the low-resolution C3K stellar library ($R\approx100$), which cannot resolve stellar absorption features at the resolution of the grating spectra. For each source, we therefore recomputed the stellar spectrum of the best-fit \texttt{Prospector} model with the high-resolution C3K library ($R=3000$ at rest-frame 3000--10000\,\AA). We then convolved the stellar spectrum with a Gaussian kernel whose width varies with wavelength as $\sigma_{\rm add}^2=\sigma_\star^2+\sigma_{\rm inst}^2-\sigma_{\rm lib}^2$. Here $\sigma_{\rm inst}$ is the instrumental resolution, $\sigma_{\rm lib}$ is the resolution of the stellar library, and $\sigma_\star$ is the stellar velocity dispersion. Since $\sigma_\star$ cannot be measured from the photometry, we estimated it at a given stellar mass and redshift using the stellar mass--velocity dispersion relation from \citet{Cannarozzo2020}, which gives $\sigma_\star\approx250\,\kms$ for massive galaxies at $z=3-5$. The smooth AGN continuum was then added and the resulting continuum template was resampled onto the observed wavelength grid of each grating spectrum. 

\subsection{Spectral decomposition and identification of broad-line AGNs}
\label{sec:specfit}

\begin{figure*}
\centering
\includegraphics[width=\hsize]{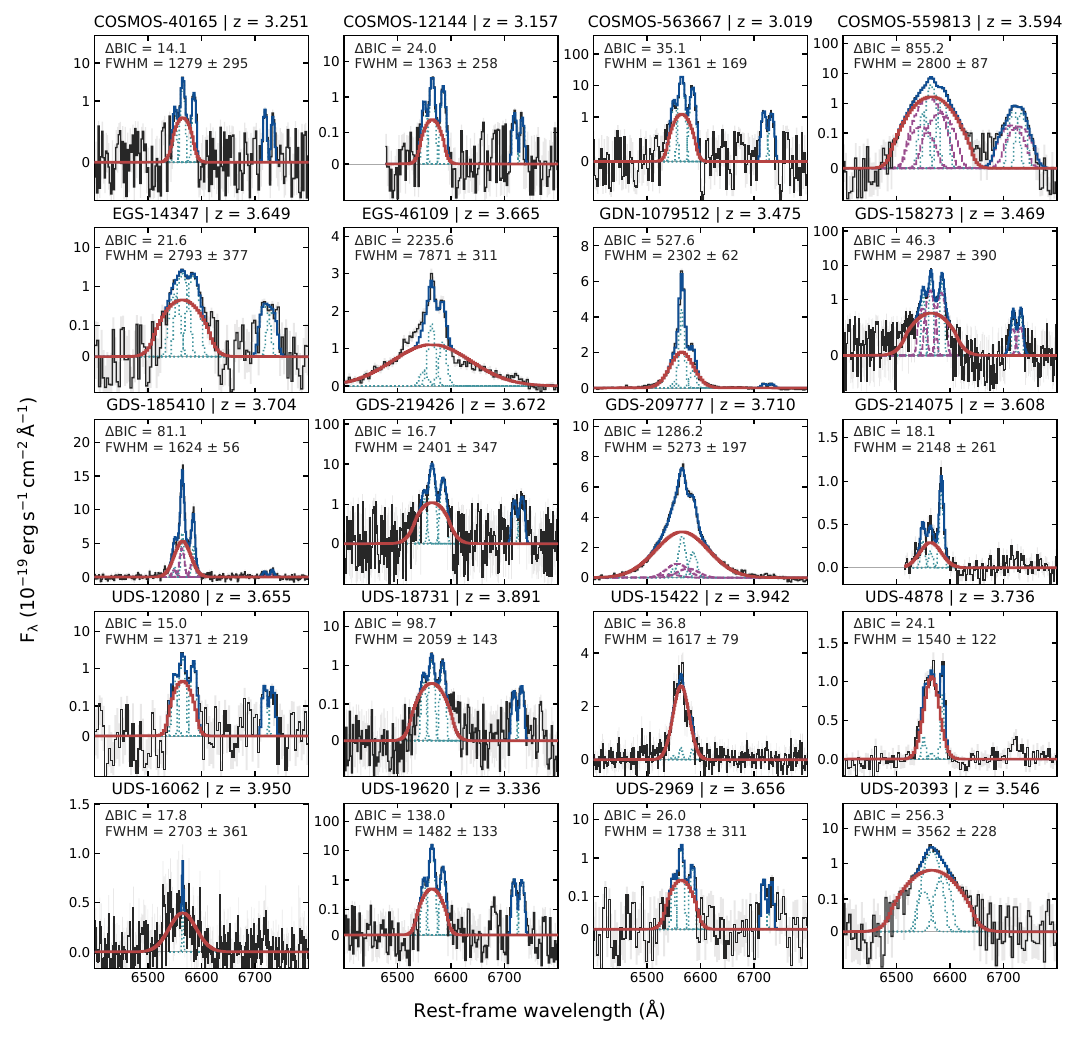}
\caption{Spectral fitting of the \ha\ region for the 20 BLAGNs in the final sample. Each panel shows the continuum-subtracted spectrum (black) and the best-fit emission-line model (blue). The individual broad-line region (BLR; red), narrow-line (dashed green), and outflow (dashed pink) components are also shown. A logarithmic flux scale is used in some panels to better illustrate the broad-line components.}
\label{fig:spec}
\end{figure*}

We searched for broad \ha\ emission in all medium- and high-resolution NIRSpec grating spectra (G140M/H, G235M/H, and G395M/H) of the massive galaxy sample that cover the \ha+\nii\ complex. Before classifying a source as a BLAGN, we assessed whether the broad emission originated from the broad-line region (BLR) or could instead be attributed to outflows \citep{Taylor2024, Maiolino2023, Harikane2023}.

We fitted the spectra using \texttt{lmfit} with the Levenberg--Marquardt optimization algorithm \citep{lmfit}. Each spectrum was fitted over rest-frame 6400--6800 \AA\ using the continuum model described in Section~\ref{sec:continuum} and Gaussian emission-line profiles. For these fits, we used spectra corrected only by the standard \texttt{msaexp} slit-loss correction, without applying the additional photometric recalibration described in Section~\ref{sec:calib}. This is because the NIRCam photometry includes substantial contributions from extended stellar populations, whereas the AGN broad lines originates from the central region. Before fitting, we rescaled the pipeline error spectrum in the line-fitting region such that its median matches the sigma-clipped scatter of the line-free continuum.

We first fitted a narrow-line-only model (model A), consisting of narrow \ha\ and the \nii\,$\lambda\lambda$6549,6585 doublet with a fixed flux ratio of 1:2.9. All narrow lines shared a common intrinsic width within $25\le \mathrm{FWHM} \le 1000$\,\kms\ and a common velocity offset relative to the DJA redshift. We restricted this offset to a wavelength shift of $\pm1$~\AA. When detected at $\mathrm{S/N}>5$, the \sii\,$\lambda\lambda$6718,6733 doublet was fitted simultaneously to provide additional constraints on the deblending of \ha\ and \nii. Model~B then adds a broad \ha\ component centered on narrow \ha, with $500\le \mathrm{FWHM}\le 15000$\,\kms. We forward-modeled each emission-line profile through the instrumental line spread function (LSF). The continuum model included a single multiplicative normalization.

We compared models using $\mathrm{BIC}=\chi^2+k\ln n$, where $k$ is the number of free parameters and $n$ is the number of fitted pixels \citep{Raftery1995}. An initial broad \ha\ detection requires $\Delta\mathrm{BIC}=\mathrm{BIC_A}-\mathrm{BIC_B}>10$, $\mathrm{FWHM}>1000$ \kms, and a width more than twice that of the narrow lines. However, outflows can also produce broad emission lines. To test this possibility, we searched for corresponding broad wings in narrow forbidden lines in \oiii\ or \sii\ (if \oiii\ was not covered or detected). We compared a single-Gaussian model with a two-component model consisting of a narrow core and a broad wing. Based on these fits, we considered three cases when assessing the origin of broad \ha\ emission:
\begin{enumerate}
\item When the core+wing model was preferred by $\Delta\mathrm{BIC}>10$, we added an outflow component to each narrow emission line in the \ha\ fitting window (model C). These outflow components share a common width and velocity offset, and cannot exceed the corresponding values derived from \oiii\ or \sii. We then added a BLR \ha\ component (model D) and classified the source as a BLAGN only if $\mathrm{BIC_C}-\mathrm{BIC_D}>10$ and the BLR component satisfies the width criteria described above.
\item When \oiii\ or \sii\ was detected without evidence of outflow signatures, we attributed the broad \ha\ component in model B to BLR emission.
\item When neither \oiii\ nor \sii\ was covered or detected, we fitted an outflow-only model (model C) with unconstrained wing parameters. We favored the BLR interpretation (model B) unless model C, which has one additional free parameter, provided a significantly better fit with $\mathrm{BIC_B}-\mathrm{BIC_C}>10$.
\end{enumerate}

This procedure identified 26 BLAGNs, 6 of which were excluded because they were classified as little red dots (LRDs) \citep[e.g.][]{Greene2024, Pan2026}. These objects were identified by their characteristic V-shaped continua in PRISM spectra (3 sources, also reported in \citealt{Hviding2025}) or broadband SEDs (3 sources). For these systems, our SED models and virial mass estimators may not yield reliable stellar and black hole masses. The physical origin of their distinctive SED shape remains debated, with proposed explanations including a dust-reddened AGN combined with a blue stellar component, scattered AGN light, or an AGN embedded in dense gas that produces a Balmer break. These different interpretations can yield stellar mass estimates that differ by orders of magnitude \citep{Wang2024LRD}. Furthermore, electron scattering rather than virial motion may be responsible for the broadening of their Balmer lines, leading to significant overestimates of black hole masses derived from virial estimators \citep{Rusakov2026}.

The final sample contains 20 BLAGNs at $3.02\le z\le 3.95$, with a median redshift of $z=3.65$. Table~\ref{tab:phys} summarizes their basic properties and Figure~\ref{fig:spec} presents their spectral fitting results. When multiple spectra satisfied the broad-line identification criteria, we adopted the measurements from the fit with the largest $\Delta\mathrm{BIC}$. Four BLAGNs were identified using model D, where an \oiii\ outflow was detected but an additional BLR component was still required. Ten sources exhibited forbidden emission lines without evidence of outflows, while the remaining six lacked detectable outflow tracers and were classified through the model B--C comparison.
The overall non-LRD BLAGN fraction of our sample is $\approx17\%$. This fraction likely represents a lower limit to the intrinsic BLAGN incidence, as broad lines from fainter or more obscured AGNs may remain undetected. 

\subsection{Black hole masses and bolometric luminosities}
\label{sec:mbh}

We estimated black hole masses using the single-epoch virial method \citep{Vestergaard2006, Shen2024RM}, which combines the BLR size inferred from the radius--luminosity relation calibrated using reverberation-mapped AGNs with the gas velocity inferred from the broad line width. We adopted the broad-\ha\ calibration of \citet{Reines2015}:
\begin{equation}\label{eq:mbh}
\begin{split}
\log\left(\frac{\mbh}{\msun}\right) = {}& 6.57 + 0.47\,\log\left(\frac{L_{\rm H\alpha,b}}{10^{42}\,\ergs}\right)\\
& + 2.06\,\log\left(\frac{\mathrm{FWHM_{H\alpha,b}}}{10^{3}\,\kms}\right),
\end{split}
\end{equation}
where $L_{\rm H\alpha,b}$ is the broad \ha\ luminosity, and $\mathrm{FWHM_{H\alpha,b}}$ is the corresponding line width. We propagated the statistical uncertainties in the fitted line flux and width and added the intrinsic scatter of the virial mass estimator (0.35 dex) in quadrature to obtain the final black hole mass uncertainties \citep{Shen2024RM}.

\begin{figure}
\centering
\includegraphics[width=\hsize]{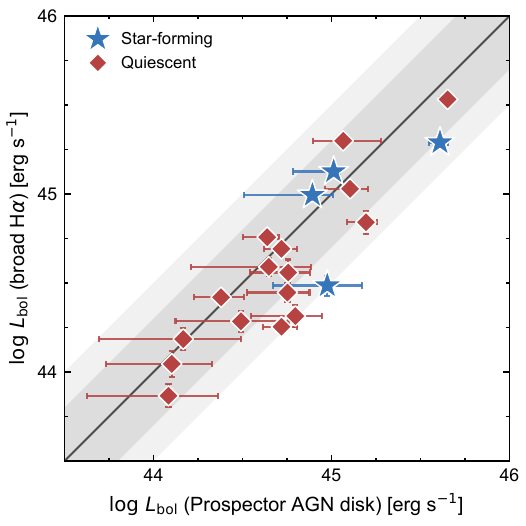}
\caption{Comparison of AGN bolometric luminosities derived from SED fitting and broad \ha\ emission. The horizontal-axis values are estimated from the rest-frame 5100\,\AA\ luminosity of the AGN accretion-disk component in the best-fit \texttt{Prospector} model, while the vertical-axis values are inferred from the broad \ha\ luminosity using the $L_{\rm H\alpha}$--$L_{5100}$ relation of \citet{Greene2005}. Stars and diamonds represent star-forming and quiescent hosts, respectively (Section~\ref{sec:hostclass}). The shaded regions indicate deviations of $\pm0.3$ and $\pm0.5$\,dex from the one-to-one relation.}
\label{fig:lbol}
\end{figure}

We estimated the AGN bolometric luminosities from broad \ha\ emission by first deriving the rest-frame 5100 \AA\ continuum luminosities using the relation of \citet{Greene2005}. We then applied a bolometric correction of 9.26 to derive \lbol\ \citep{Richards2006}. As an independent estimate, we also derived the 5100\,\AA\ continuum luminosities from the AGN accretion-disk component of the best-fit SED models. The two estimates are in good agreement, with a median offset of $-0.10$\,dex and a scatter of 0.22\,dex (Figure~\ref{fig:lbol}). This agreement supports the consistency between our spectroscopic and SED decomposition results.

\section{Results}
\label{sec:results}

\subsection{Moderate-luminosity AGNs in massive, evolved hosts}
\label{sec:properties}

The broad \ha\ lines of the final BLAGN sample have FWHM of 1300--7900~\kms\ (median 2100~\kms) and luminosities of $\log\,L_{\rm H\alpha,b}/\ergs=41.4-43.4$ (median 42.2). The corresponding black hole masses span $\logmbh=6.6-8.9$ (median 7.5), while the bolometric luminosities span $\loglbol=43.9-45.5$ (median 44.6). The Eddington ratio ranges from 0.02 to 0.48, with a median of 0.13. In terms of their SMBH properties, our AGNs are thus similar to the bulk of faint BLAGNs identified in deep JWST surveys at $z\sim4$--7 \citep[e.g.][]{Harikane2023, Maiolino2023, Juodzbalis2026census}, but are more than two orders of magnitude fainter than the $z\sim6$ quasars with host-galaxy measurements \citep{Ding2023, Stone2024, Yue2024}. The host galaxies of our AGNs, however, are 1--3 orders of magnitude more massive than those of most faint JWST AGNs discovered in individual pencil-beam surveys \citep{Harikane2023, Juodzbalis2026census}. 

Two sources are noteworthy. EGS-46109 exhibits an exceptionally broad \ha\ line ($\mathrm{FWHM}\approx7900$ \kms) with $\logmbh \approx 8.92$, and shows an AGN-dominated SED ($f_{\rm AGN} \approx 0.84$), with a mass ratio ($\mbh/\m\approx8\%$) comparable to those of overmassive quasars at $z\sim6$. GDS-209777, a quiescent galaxy with $\logm \approx 10.12$, also hosts an overmassive black hole ($\logmbh \approx 8.69$) with $\mbh/\m\approx3.8\%$, and lies about 1 dex above the mass relation of local early-type galaxies \citep{Reines2015}.

The remaining AGNs contribute only modestly to the observed continuum, allowing their stellar masses and star formation rates to be more reliably constrained than those of AGN-dominated systems. Many sources exhibit a prominent Balmer break and a red rest-frame optical continuum in their SEDs (Fig.~\ref{fig:sed}) and PRISM spectra. This contrasts with the majority of low-mass AGNs identified in individual pencil-beam surveys, whose blue UV--optical continua are often dominated by the central AGN \citep{Maiolino2023}. The narrow \nii/\ha\ flux ratio of our BLAGNs has a median of 0.6, several times higher than that of low-stellar-mass AGNs at $z>4$ \citep{Harikane2023, Maiolino2023}. This elevated ratio is consistent with our AGNs residing in massive, chemically enriched galaxies.

\subsection{Star-forming and quiescent hosts}
\label{sec:hostclass}

We classified the host galaxies based on their specific star formation rates (sSFRs), defined as the SFR averaged over the past 100\,Myr divided by the stellar mass. Following \citet{Carnall2023}, we classified galaxies as quiescent if $\mathrm{sSFR}<0.2/t_{\rm H}(z)$, where $t_{\rm H}(z)$ is the age of the Universe at the source redshift. At the median redshift of our sample, this criterion corresponds to a threshold of $\log\,\mathrm{sSFR}/\mathrm{yr}^{-1}\approx-10$. Based on this criterion, 16 hosts were classified as quiescent and 4 as star-forming (Table~\ref{tab:phys}).

However, the inferred sSFRs have substantial uncertainties. To assess the robustness of these classifications, we computed the Bayes factor for each source as $\mathrm{BF}=[P_{\rm post}/(1-P_{\rm post})]/[P_{\rm prior}/(1-P_{\rm prior})]$, where $P_{\rm post}$ is the fraction of posterior samples below the quiescence threshold and $P_{\rm prior}$ is the corresponding fraction obtained by sampling the model prior at the fitted stellar mass. The Bayes factor thus quantifies how strongly the photometric data shift the odds of quiescence relative to the SFH prior with $P_{\rm prior}\approx0.5$. Adopting $\mathrm{BF}>3$ and $\mathrm{BF}<1/3$ as evidence for quiescence and star formation, respectively, we find that 11 of the 16 quiescent hosts and 2 of the 4 star-forming hosts meet these criteria, while the remaining 7 sources lack sufficient evidence for either classification. With these uncertainties in mind, we find that the quiescent hosts are slightly more massive, with a median $\logm\approx10.64$, compared to 10.38 for the star-forming hosts.

The high quiescent fraction may partly reflect selection biases in the parent sample, as several NIRSpec programs, including RUBIES \citep{deGraaff2024} and EXCELS \citep{Carnall2024}, prioritized red galaxies in their target selection or specifically targeted quiescent systems. In addition, faint broad emission lines may be more easily detected against the continuum of quiescent galaxies with low SFRs. We therefore caution that the observed quiescent fraction may not be representative of the overall population of massive BLAGN hosts.

\begin{figure*}
\centering
\includegraphics[width=\hsize]{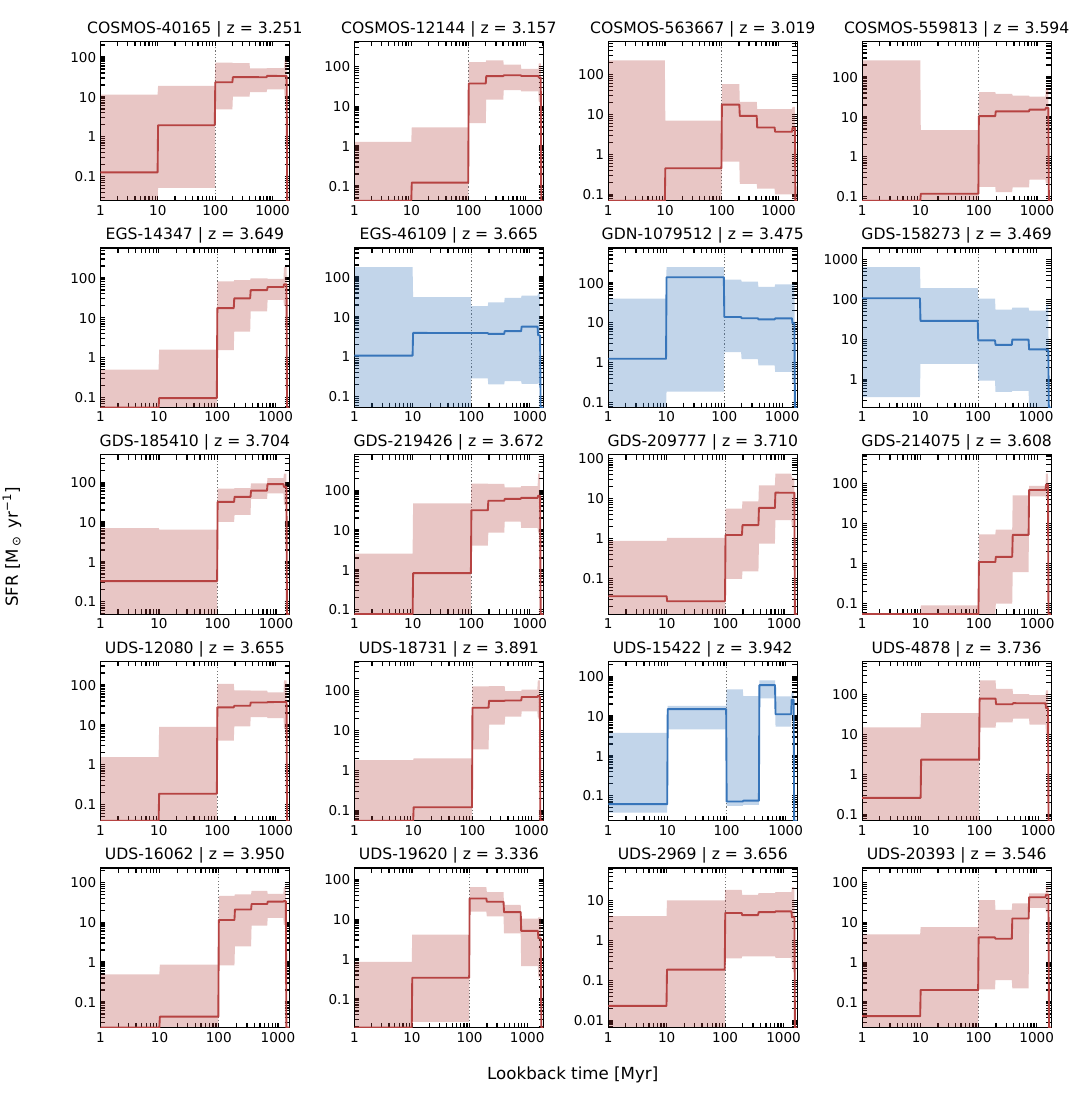}
\caption{Reconstructed star formation histories and their uncertainties for the 20 BLAGNs in the final sample. Star-forming and quiescent hosts are shown in blue and red, respectively.}\label{fig:sfh}
\end{figure*}

Although the classification of these sources as fully quiescent remains uncertain because their sSFR uncertainties overlap the adopted threshold, the evidence for recent quenching is more robust, as indicated by their pronounced Balmer breaks and weak rest-frame UV emission, resembling those of post-starburst galaxies (Figure~\ref{fig:sed}). The reconstructed SFHs of our quiescent AGN hosts suggest that they formed most of their stellar mass during an early, intense star-forming episode, followed by a rapid decline in star formation within the past few hundred Myrs (Figure~\ref{fig:sfh}). These characteristics are consistent with the growing population of spectroscopically confirmed massive quiescent galaxies at $3<z<5$ \citep[e.g.,][]{Carnall2023nat, Carnall2024, Glazebrook2024, Nanayakkara2024, deGraaff2024qs}.

\subsection{The \texorpdfstring{\rel}{MBH-M*} relation}
\label{sec:massrel} 
 
\begin{figure} 
\centering
\includegraphics[width=\hsize]{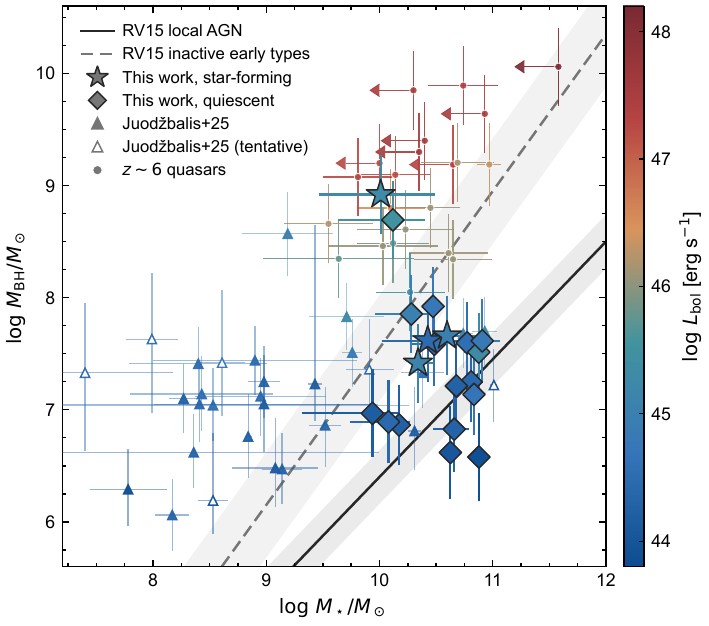}
\caption{Black hole mass versus host-galaxy stellar mass for high-redshift AGNs. Large stars and diamonds represent the star-forming and quiescent AGNs in this work, respectively. Triangles represent the JADES BLAGNs at $z\sim2-9$ from \citet[][open symbols: tentative detections]{Juodzbalis2026census}, while circles represent $z\sim6$ quasars with JWST stellar mass measurements \citep{Ding2023, Stone2024, Yue2024}, with arrows indicating upper limits on stellar mass. All data points are color-coded by \lbol. The solid and dashed lines show the local relations for AGNs and inactive early-type galaxies from \citet{Reines2015}, with the shaded regions indicating their intrinsic scatter.}
\label{fig:MM}
\end{figure}

Figure~\ref{fig:MM} compares our BLAGNs with local galaxies and previously reported high-redshift AGNs in the \rel\ plane. We show the local relations for BLAGNs with virial BH mass estimates and inactive early-type galaxies with dynamical black hole mass measurements \citep{Reines2015}. The latter have $\sim$1 dex higher $\mbh/\m$ than the predominantly star-forming, disk-dominated hosts of local AGNs \citep{Reines2015, Greene2020}. Although most of our hosts are quiescent, their SMBHs are distributed around the local AGN relation, with typical $\mbh/\m\sim0.1\%$. They lie well below the local early-type relation, in contrast to many previously reported faint JWST AGNs and high-redshift quasars with $\mbh/\m\sim10\%$ \citep{Maiolino2023, Harikane2023, Stone2024, Yue2024}. Our sample thus represents the long-sought population of normal-mass SMBHs in the distant Universe.

These differences are consistent with the selection effects discussed in Section~\ref{sec:intro}. Shallow quasar surveys miss fainter AGNs powered by lower-mass black holes, including those in massive host galaxies. Deep pencil-beam JWST surveys can detect AGNs with \lbol\ as low as $10^{44}\ \ergs$ at $z>3$, but their detected populations are preferentially biased toward overmassive black holes in the more numerous low-mass galaxies due to the Lauer bias. Indeed, the small number of JWST AGNs previously identified in massive host galaxies already lie close to the mean local relation \citep{ Juodzbalis2026census, SunY2025}, as JWST is sufficiently sensitive to detect normal-mass black holes in these galaxies. With our dedicated sample selection, the statistics in the high-$\m$, low-$\mbh$ regime are now robust.  Notably, our sample is consistent with the intrinsic high-redshift mass relations inferred by \cite{Li2024} and \cite{Silverman2025}, who used forward modeling to account for selection biases in samples of overmassive AGNs and quasars. Nevertheless, even lower-mass black holes with fainter luminosities ($\lbol < 10^{44}\ \ergs$) or narrower emission lines ($\rm FWHM < 1000\ \kms$) may still be missed in our search.

In addition, we find that quiescent hosts do not harbor more massive black holes relative to their stellar masses than star-forming hosts (Figure~\ref{fig:MM}). Their median \mbh/\m\ ratio is 0.06\%, lower than the 0.24\% found for star-forming hosts, although the small sample size limits the statistical significance of this difference. This finding challenges the theoretical expectation that galaxy quenching requires an overmassive black hole. We discuss its implications in Section~\ref{sec:quenching}.

\begin{table*}
\caption{Information of the 20 BLAGNs in the final sample.}
\label{tab:obs}
\centering
\begin{tabular}{lcccccc}
\hline\hline
ID & RA & Dec & $z$ & Program & Grating & Photometry \\
 & (deg) & (deg) & & & & \\
\hline
COSMOS-40165 & 149.903602 & $2.102694$ & 3.251 & 3567 & G235M & COSMOS2025 \\   
COSMOS-12144 & 150.076370 & $2.362323$ & 3.157 & 3567 & G235M & PRIMER \\   
COSMOS-563667 & 150.465067 & $2.466406$ & 3.019 & 4318 & G235M & COSMOS2025 \\   
COSMOS-559813 & 150.486189 & $2.427975$ & 3.594 & 4318 & G395M & COSMOS2025 \\   
EGS-14347 & 214.850571 & $52.866028$ & 3.649 & 4233 & G395M & CEERS \\   
EGS-46109 & 215.079254 & $52.934258$ & 3.665 & 4233 & G395M & CEERS \\   
GDN-1079512 & 189.164840 & $62.208473$ & 3.475 & 4762 & G395M & JADES \\   
GDS-158273 & 53.072850 & $-27.908500$ & 3.469 & 1286 & G395H & JADES \\   
GDS-185410 & 53.124400 & $-27.851692$ & 3.704 & 1212 & G395H & JADES \\   
GDS-219426 & 53.158386 & $-27.733599$ & 3.672 & 1212 & G395H & JADES \\   
GDS-209777 & 53.158466 & $-27.774048$ & 3.710 & 1180 & G235M & JADES \\   
GDS-214075 & 53.196910 & $-27.760532$ & 3.608 & 1207 & G235M & JADES \\   
UDS-12080 & 34.238393 & $-5.205769$ & 3.655 & 4233 & G395M & PRIMER \\   
UDS-18731 & 34.274451 & $-5.143820$ & 3.891 & 3543 & G235M & PRIMER \\   
UDS-15422 & 34.322559 & $-5.171390$ & 3.942 & 1215 & G395H & PRIMER \\   
UDS-4878 & 34.330968 & $-5.287489$ & 3.736 & 4233 & G395M & PRIMER \\   
UDS-16062 & 34.338211 & $-5.166201$ & 3.950 & 1215 & G395H & PRIMER \\   
UDS-19620 & 34.386040 & $-5.135954$ & 3.336 & 3543 & G235M & PRIMER \\   
UDS-2969 & 34.405873 & $-5.312953$ & 3.656 & 4233 & G395M & PRIMER \\   
UDS-20393 & 34.410784 & $-5.129682$ & 3.546 & 4233 & G395M & PRIMER \\   
\hline
\end{tabular}
\tablefoot{The identifier combines the field (GDS: GOODS-S; GDN: GOODS-N) with the source ID in the photometric catalog listed in the last column. RA and Dec are the source positions in the photometric catalog. $z$ is the DJA spectroscopic redshift. Program and grating refer to the spectrum from which the broad-line measurements are taken.}
\end{table*}

\begin{table*}
\caption{Broad-line, black hole and host-galaxy properties of the final BLAGN sample.}
\label{tab:phys}
\centering
\small
\setlength{\tabcolsep}{3pt}
\begin{tabular}{lcccccccccccc}
\hline\hline
ID & FWHM$_{\rm H\alpha,b}$ & $\log L_{\rm H\alpha,b}$ & S/N & $\Delta$BIC & Model & \nii/\ha & $\log\mbh$ & $\log\m$ & $f_{\rm AGN}$ & $\log\mathrm{sSFR}$ & Class & $\log L_{\rm bol}$ \\
 & (\kms) & (\ergs) & & & & & (\msun) & (\msun) & & (yr$^{-1}$) & & (\ergs) \\
\hline
COSMOS-40165 & $1279\pm295$ & $41.63\pm0.09$ & 5.0 & 14.1 & B & 0.41 & $6.62\pm0.41$ & $10.63^{+0.07}_{-0.08}$ & 0.08 & $-10.23$ & Q & 44.05 \\
COSMOS-12144 & $1363\pm258$ & $41.43\pm0.07$ & 5.8 & 24.0 & B & 0.56 & $6.58\pm0.40$ & $10.88^{+0.05}_{-0.06}$ & 0.11 & $-11.56$ & Q & 43.87 \\
COSMOS-563667 & $1361\pm169$ & $42.10\pm0.06$ & 7.1 & 35.1 & B & 0.44 & $6.89\pm0.37$ & $10.08^{+0.19}_{-0.34}$ & 0.33 & $-10.49$ & Q & 44.45 \\
COSMOS-559813 & $2800\pm87$ & $42.77\pm0.04$ & 11.0 & 855.2 & D & 0.31 & $7.85\pm0.35$ & $10.28^{+0.22}_{-0.32}$ & 0.50 & $-11.26$ & Q & 45.03 \\
EGS-14347 & $2793\pm377$ & $42.23\pm0.08$ & 5.1 & 21.6 & B & 0.86 & $7.60\pm0.36$ & $10.77^{+0.08}_{-0.08}$ & 0.35 & $-11.66$ & Q & 44.56 \\
EGS-46109 & $7871\pm311$ & $43.07\pm0.02$ & 23.1 & 2235.6 & B & 0.72 & $8.92\pm0.35$ & $10.01^{+0.48}_{-0.55}$ & 0.84 & $-8.93$ & SF & 45.29 \\
GDN-1079512 & $2302\pm62$ & $42.73\pm0.02$ & 24.7 & 527.6 & B & 0.24 & $7.66\pm0.35$ & $10.60^{+0.38}_{-0.33}$ & 0.38 & $-8.40$ & SF & 44.99 \\
GDS-158273 & $2987\pm390$ & $42.14\pm0.07$ & 6.7 & 46.3 & D & 0.85 & $7.62\pm0.37$ & $10.43^{+0.43}_{-0.41}$ & 0.29 & $-8.30$ & SF & 44.49 \\
GDS-185410 & $1624\pm56$ & $43.08\pm0.04$ & 12.1 & 81.1 & D & 0.68 & $7.51\pm0.35$ & $10.88^{+0.06}_{-0.13}$ & 0.17 & $-10.84$ & Q & 45.30 \\
GDS-219426 & $2401\pm347$ & $42.56\pm0.07$ & 5.8 & 16.7 & B & 0.42 & $7.61\pm0.37$ & $10.91^{+0.16}_{-0.29}$ & 0.65 & $-10.93$ & Q & 44.84 \\
GDS-209777 & $5273\pm197$ & $43.35\pm0.03$ & 16.7 & 1286.2 & D & 0.65 & $8.69\pm0.35$ & $10.12^{+0.29}_{-0.48}$ & 0.89 & $-11.49$ & Q & 45.53 \\
GDS-214075 & $2148\pm261$ & $41.91\pm0.07$ & 6.3 & 18.1 & B & -- & $7.21\pm0.37$ & $10.68^{+0.06}_{-0.05}$ & 0.13 & $-12.86$ & Q & 44.28 \\
UDS-12080 & $1371\pm219$ & $41.95\pm0.07$ & 6.0 & 15.0 & B & 0.67 & $6.83\pm0.38$ & $10.66^{+0.13}_{-0.19}$ & 0.41 & $-11.22$ & Q & 44.31 \\
UDS-18731 & $2059\pm143$ & $42.07\pm0.04$ & 12.4 & 98.7 & B & 0.70 & $7.25\pm0.36$ & $10.81^{+0.06}_{-0.06}$ & 0.18 & $-11.52$ & Q & 44.42 \\
UDS-15422 & $1617\pm79$ & $42.89\pm0.03$ & 17.2 & 36.8 & B & 0.94 & $7.42\pm0.35$ & $10.34^{+0.14}_{-0.03}$ & 0.31 & $-9.20$ & SF & 45.13 \\
UDS-4878 & $1540\pm122$ & $42.38\pm0.03$ & 12.8 & 24.1 & B & -- & $7.14\pm0.36$ & $10.83^{+0.09}_{-0.09}$ & 0.29 & $-10.29$ & Q & 44.69 \\
UDS-16062 & $2703\pm361$ & $42.27\pm0.06$ & 7.7 & 17.8 & B & -- & $7.58\pm0.38$ & $10.48^{+0.11}_{-0.20}$ & 0.29 & $-11.63$ & Q & 44.59 \\
UDS-19620 & $1482\pm133$ & $41.88\pm0.04$ & 12.3 & 138.0 & B & 0.18 & $6.86\pm0.36$ & $10.17^{+0.06}_{-0.07}$ & 0.25 & $-10.63$ & Q & 44.25 \\
UDS-2969 & $1738\pm311$ & $41.80\pm0.07$ & 5.9 & 26.0 & B & 0.28 & $6.97\pm0.39$ & $9.95^{+0.24}_{-0.62}$ & 0.37 & $-10.24$ & Q & 44.18 \\
UDS-20393 & $3562\pm228$ & $42.46\pm0.03$ & 13.1 & 256.3 & B & 0.20 & $7.92\pm0.35$ & $10.47^{+0.04}_{-0.05}$ & 0.25 & $-10.74$ & Q & 44.76 \\
\hline
\end{tabular}
\tablefoot{FWHM$_{\rm H\alpha,b}$ is the line width of the broad \ha\ component, and $L_{\rm H\alpha,b}$ is its luminosity, S/N indicates the significance of its detection. Model B uses narrow lines plus a broad \ha\ component, Model D adds an additional outflow component in each narrow line. $\Delta$BIC is the improvement of the selected model over the same model without the BLR \ha\ component. \nii/\ha\ is the narrow-line flux ratio in the selected model. $f_{\rm AGN}$ is the AGN fraction at rest-frame 0.51\,$\mu$m in the SED fit. sSFR is the ratio of the SFR averaged over 100~Myrs to the stellar mass. Class: Q, quiescent; SF, star-forming. $L_{\rm bol}$ is derived from $L_{\rm H\alpha,b}$ \citep{Greene2005, Richards2006}.}
\end{table*}

\section{Discussion}
\label{sec:discussion}

\subsection{Back-tracing the mass ratio to \texorpdfstring{$z=6$}{z=6}}
\label{sec:z6}

\begin{figure*}
\centering
\includegraphics[width=\hsize]{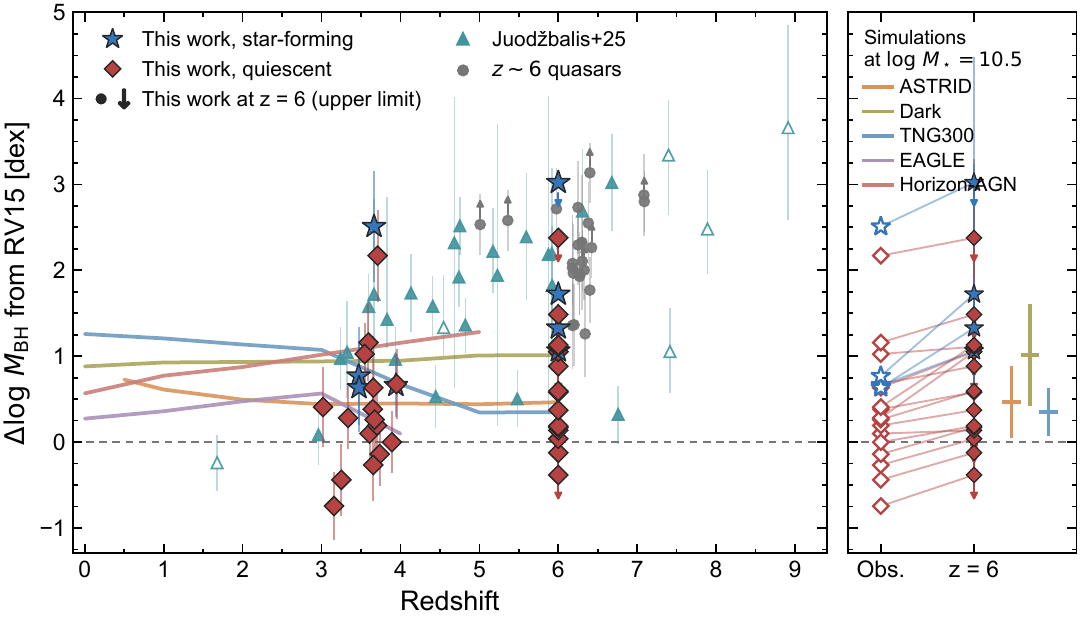}
\caption{Black hole mass offsets relative to the local AGN relation of \citet{Reines2015}, evaluated at the stellar mass of each source. Left: Offsets as a function of redshift for the AGNs in this work (blue stars: star-forming; red diamonds: quiescent), the JADES BLAGNs from \citet[][triangles; open symbols represent tentative detections]{Juodzbalis2026census}, and $z\sim6$ quasars \citep[][gray circles]{Ding2023, Stone2024, Yue2024}. The solid curves show the mean offsets predicted at $\logm=10.5$ by the ASTRID, Dark Sage, TNG300, EAGLE, and Horizon-AGN models \citep{Habouzit2021, Ni2024, Dattathri2024}. Filled blue stars and red diamonds with downward arrows at $z=6$ indicate the reconstructed upper limits for our BLAGNs. Right: For each of our BLAGNs, the observed offset (open symbol) is connected to its reconstructed upper limit at $z=6$ (filled symbol) by a solid line. Colored bars indicate the mean and dispersion of the simulation predictions at $z=6$.}
\label{fig:deltambh}
\end{figure*}

The SED fits allow us to reconstruct where our galaxies were in the \mbh-\m\ plane back to the epoch of reionization and investigate the possible evolutionary pathways of their $\mbh/\m$ ratios during the first billion years of cosmic history. We integrated the best-fit SFH up to the cosmic time corresponding to $z=6$ to estimate the stellar mass assembled at that epoch. For a galaxy observed at $t_{\rm obs}$ with a stellar age $T$, the population at $z=6$ has an age of $T_6 = T - (t_{\rm obs} - t_6)$, where $t_6$ is the cosmic time at $z=6$. We accounted for stellar mass loss by weighting the mass formed at earlier times by the age-dependent fraction remaining in stars and stellar remnants ($f_{\rm ret}$) computed by FSPS:
\begin{equation}
    \m(z=6) = \int_{0}^{T_6} {\rm SFR}(u)\, f_{\rm ret}(T_6 - u)\, {\rm d}u,
\end{equation}
where $u$ is the time since the onset of star formation. Given that black hole masses do not decrease over time, the observed \mbh\ thus provides an upper limit on the black hole mass at $z=6$ and, consequently, on the corresponding \mbh/\m\ ratio.

As shown in Fig.~\ref{fig:deltambh}, the reconstructed mass relation at $z\sim6$ suggests that some progenitors of our massive galaxies could have hosted normal-mass or even undermassive black holes, with \mbh/\m\ ratios comparable to or below the local AGN relation in the reionization epoch. This points to a potentially important population of early SMBHs that remains largely unexplored by current high-redshift AGN surveys.

Several observational effects may hinder the identification of such systems. At $z>6$, their host galaxies were likely undergoing active star formation, producing blue spectral continua that can resemble those of unobscured AGNs \citep{Maiolino2023, Boyett2024}. Stellar mass estimates also become more uncertain when the available NIRCam photometry probe only blue wavelengths, while the outshining of older stellar populations by young stars can lead to underestimated stellar masses \citep{Gimenez2023, Arteaga2024}. Moreover, identifying BLR emission may be more challenging at $z>6$, as the lower black hole masses expected for these progenitors could produce narrower broad-line profiles, potentially overlapping with those produced by starburst-driven outflows. Nevertheless, our reconstructed mass relation suggests that such systems may constitute a significant population at $z>6$, and recent spatially resolved NIRSpec IFU studies have begun to uncover potential systems \citep{Ren2025}.

The coexistence of these systems with other AGNs and quasars hosting overmassive black holes across a wide range of masses points to diverse SMBH formation and growth histories in the early Universe. The most overmassive black holes may have originated from heavy seeds or undergone episodes of rapid, potentially super-Eddington accretion \citep{Natarajan2024, Trinca2026}. Strong AGN feedback in some of these systems may have rapidly quenched their compact host galaxies \citep{Onoue2024}, which could subsequently evolve into the dense cores of present-day massive elliptical galaxies \citep{Baggen2023}. Their BHs may remain overmassive even after substantial host growth through mergers. In contrast, the relatively normal-mass black holes in our massive galaxies are consistent with an alternative evolutionary pathway involving lighter seeds and delayed BH growth. In this scenario, supernova feedback initially suppresses black hole accretion by removing gas from the central regions of low-mass galaxies. Efficient black hole growth begins only after the host develops a sufficiently deep gravitational potential to retain cold gas and sustain nuclear accretion \citep{Habouzit2017, Greene2020, Zhu2022}. Alternatively, black holes may have grown in lockstep with the stellar mass growth of their hosts, maintaining relatively constant \mbh/\m\ ratios throughout their evolution while reducing the scatter in the mass relation through mergers \citep{Tanaka2026}.

Although our reconstruction cannot uniquely distinguish between these scenarios because the accretion history of individual black holes cannot be directly traced, it highlights the possibility that the progenitors of $z\sim3$ massive galaxies may have hosted a substantial, yet largely unexplored population of low-mass black holes during the epoch of reionization. Overall, the observed diversity in the \mbh/\m\ ratios suggests that the present-day black hole–galaxy scaling relations may have emerged through multiple seeding and growth pathways, rather than a single heavy seed formation channel inferred from biased samples dominated by overmassive black holes.

\subsection{Comparison with cosmological simulations}
\label{sec:sims}

Fig.~\ref{fig:deltambh} compares the offsets of our sample from the local AGN mass relation with predictions from four cosmological simulations (ASTRID, TNG300, Horizon-AGN, and EAGLE) and the Dark Sage semi-analytic model, evaluated at $\logm\sim10.5$ \citep{Ni2024, Habouzit2021, Dattathri2024}. Previously reported JWST AGN \citep{Juodzbalis2026census} and quasar \citep{Ding2023, Yue2024, Stone2025} samples are also shown for comparison. Quasars, which represent extreme outliers from the underlying black hole population, are in tension with the predictions of all models. In contrast, the normal-mass black holes identified in our study are consistent with most of these models, which predict much weaker evolution in $\mbh/\m$ over cosmic time. 

Given the complex target selection criteria of our parent galaxy sample, we cannot robustly constrain the mean and intrinsic scatter of the \mbh-\m\ relation or distinguish among model predictions. Nevertheless, this comparison demonstrates how selection effects can affect the intrinsic mass relation and mislead assessments of theoretical models. On the other hand, the wide range of \mbh/\m\ ratios observed across different samples suggests that the intrinsic scatter predicted by these simulations (typically $0.2$--$0.5$ dex) is likely underestimated. Spectroscopic surveys with well-defined selection functions, such as those targeting mass-complete galaxy samples, will be essential for constraining the intrinsic mass relation and providing further insight into the physics of black hole seeding and AGN feedback.

\subsection{Implications for the quenching of massive galaxies}
\label{sec:quenching}

Despite producing a slowly evolving mean $\mbh/\m$ ratio consistent with our observations, these cosmological simulations often struggle to reproduce the observed abundance of massive quiescent galaxies at high redshift \citep{Weller2024, deGraaff2024qs}. In simulations that successfully produce such galaxies, quenching is often driven by AGN feedback from overmassive black holes \citep{Weller2024, Dattathri2024}. In contrast, the black holes in our quiescent AGN hosts have typical masses of $\logmbh\sim7.2$ and $\mbh/\m\sim0.1\%$, comparable to those in star-forming hosts. Their mass ratios also lie well below the local relation for early-type galaxies.

Our findings are in tension with assumptions and predictions from several theoretical models of AGN-driven quenching. For instance, in the ASTRID simulation, the kinetic feedback mode of low-accretion-rate AGNs, which suppresses gas cooling in the circumgalactic medium and is primarily responsible for quenching massive galaxies, is activated only for SMBHs with $\mbh>10^{8.5}\,\msun$ at $z<2.3$. In contrast, the thermal feedback mode operating at $z>2.3$ is insufficient to effectively quench massive galaxies \citep{Ni2024, Dattathri2024}. Similarly, the IllustrisTNG simulation predicts that massive quiescent galaxies at $z\sim3$ exclusively host overmassive black holes with $\logmbh>8.5$ \citep{Weller2024}. The analytic model of \citet{Pacucci2024} further links galaxy quenching to the mass ratio. However, at the median Eddington ratio of our sample ($\edd\sim0.1$), the model predicts a quenching threshold of $\mbh/\m\sim1\%$ for massive high-redshift galaxies, more than an order of magnitude higher than the median ratio observed in our quiescent hosts ($\mbh/\m\approx0.06\%$).

Reconciling cosmological simulations with recent JWST observations of massive quiescent galaxies at high redshift may require more efficient quenching mechanisms that do not depend on the presence of overmassive black holes. Possible mechanisms include stronger supernova feedback and more effective starburst-driven outflows than those currently implemented in cosmological simulations \citep{Deng2024}, as well as more efficient AGN feedback from lower-mass black holes.

After quenching, mass loss from evolved stars may provide an important source of fuel for black hole accretion. This continued growth, together with mergers involving galaxies hosting overmassive black holes, may gradually establish the $\mbh-\m$ relation observed in local elliptical galaxies \citep{Kormendy2013, Aird2022}. The supply of recycled stellar gas could also explain the prevalence of AGN activity in quiescent galaxies \citep{Aird2022, Baker2024}. As these black holes grow more massive, radio-mode AGN feedback may eventually become effective in maintaining long-term quiescence in their host galaxies. Future IFU observations probing stellar populations, kinematics, outflows, and residual star formation, complemented by measurements of radio activity and gas reservoirs of our sample, will provide valuable insights into the evolutionary histories of these galaxies and help constrain the next generation of high-resolution cosmological simulations.

\section{Conclusions}
\label{sec:conclusions} 
We conducted a systematic search for broad-line AGNs in massive galaxies at $3<z<5$ using archival JWST/NIRSpec spectroscopy from DJA v4.4. Among 117 massive galaxies with $\logm \gtrsim 10$ and usable \ha\ spectra, we identified 20 non-LRD AGNs with broad \ha\ detected at $\mathrm{S/N}>5$. We characterized their black holes and host-galaxy properties through multiband SED fitting and spectral decomposition.

Our AGNs have moderate luminosities ($\loglbol\sim44.5$), black hole masses ($\logmbh\sim7.5$) and Eddington ratios ($\edd\sim0.1$), comparable to those of faint JWST AGNs reported in previous studies. However, our host galaxies are substantially more massive ($\logm \approx 10.0-11.0$), owing to the large combined survey area that enables the identification of these rare massive systems.

The median black hole-to-stellar mass ratio of our sample is $\mbh/\m\approx0.1\%$, consistent with that observed in local AGNs. Our results demonstrate that SMBHs at high redshift are not universally overmassive relative to their hosts. Instead, the elevated mass ratios commonly reported at high redshift may arise, at least in part, from selection effects that preferentially identify the upper envelope of a broad underlying black hole mass distribution. The coexistence of overmassive and normal-mass black holes in the early Universe suggests diverse black hole seeding and evolutionary pathways.

Furthermore, we find no significant difference in $\mbh/\m$ between quiescent and star-forming hosts. This suggests that the early quenching of massive galaxies does not necessarily require an overmassive black hole, challenging the strong connection between black hole mass and galaxy quenching predicted by some theoretical models. Although our findings do not rule out AGN feedback as a quenching mechanism, they highlight the need to reassess how black hole growth and feedback regulate star formation in the early Universe.

Overall, our study provides new constraints on the early growth of SMBHs and their coevolution with massive galaxies, revealing a more diverse black hole population than previously inferred from individual pencil-beam surveys. Future spectroscopic surveys with well-characterized selection functions will be essential for determining the intrinsic normalization and scatter of the $\mbh$--$\m$ relation and establishing the role of AGN feedback in the quenching of massive galaxies.

\begin{acknowledgements}
This work is based on observations obtained with the NASA/ESA/CSA James Webb Space Telescope (JWST). The data were obtained from the Mikulski Archive for Space Telescopes (MAST) at the Space Telescope Science Institute (STScI), operated by the Association of Universities for Research in Astronomy, Inc., under NASA contract NAS 5-03127 for JWST. This research used data products from the DAWN JWST Archive (DJA), an initiative of the Cosmic Dawn Center (DAWN), supported by the Danish National Research Foundation through grant DNRF140. 
\end{acknowledgements}

\bibliographystyle{aa}
\bibliography{ref}

@ARTICLE{Yue2024,
       author = {{Yue}, Minghao and {Eilers}, Anna-Christina and {Simcoe}, Robert A. and {Mackenzie}, Ruari and {Matthee}, Jorryt and {Kashino}, Daichi and {Bordoloi}, Rongmon and {Lilly}, Simon J. and {Naidu}, Rohan P.},
        title = "{EIGER. V. Characterizing the Host Galaxies of Luminous Quasars at z {\ensuremath{\gtrsim}} 6}",
      journal = {\apj},
         year = 2024,
        month = may,
       volume = {966},
       number = {2},
          eid = {176},
        pages = {176},
          doi = {10.3847/1538-4357/ad3914},
archivePrefix = {arXiv},
       eprint = {2309.04614},
 primaryClass = {astro-ph.GA},
       adsurl = {https://ui.adsabs.harvard.edu/abs/2024ApJ...966..176Y}
}

@ARTICLE{Johnson2021,
       author = {{Johnson}, Benjamin D. and {Leja}, Joel and {Conroy}, Charlie and {Speagle}, Joshua S.},
        title = "{Stellar Population Inference with Prospector}",
      journal = {\apjs},
         year = 2021,
        month = jun,
       volume = {254},
       number = {2},
          eid = {22},
        pages = {22},
          doi = {10.3847/1538-4365/abef67},
archivePrefix = {arXiv},
       eprint = {2012.01426},
 primaryClass = {astro-ph.GA},
       adsurl = {https://ui.adsabs.harvard.edu/abs/2021ApJS..254...22J}
}

@ARTICLE{Conroy2009,
       author = {{Conroy}, Charlie and {Gunn}, James E. and {White}, Martin},
        title = "{The Propagation of Uncertainties in Stellar Population Synthesis Modeling. I. The Relevance of Uncertain Aspects of Stellar Evolution and the Initial Mass Function to the Derived Physical Properties of Galaxies}",
      journal = {\apj},
         year = 2009,
        month = jul,
       volume = {699},
       number = {1},
        pages = {486-506},
          doi = {10.1088/0004-637X/699/1/486},
archivePrefix = {arXiv},
       eprint = {0809.4261},
 primaryClass = {astro-ph},
       adsurl = {https://ui.adsabs.harvard.edu/abs/2009ApJ...699..486C}
}

@ARTICLE{Conroy2010,
       author = {{Conroy}, Charlie and {Gunn}, James E.},
        title = "{The Propagation of Uncertainties in Stellar Population Synthesis Modeling. III. Model Calibration, Comparison, and Evaluation}",
      journal = {\apj},
         year = 2010,
        month = apr,
       volume = {712},
       number = {2},
        pages = {833-857},
          doi = {10.1088/0004-637X/712/2/833},
archivePrefix = {arXiv},
       eprint = {0911.3151},
 primaryClass = {astro-ph.CO},
       adsurl = {https://ui.adsabs.harvard.edu/abs/2010ApJ...712..833C}
}

@ARTICLE{Nenkova2008a,
       author = {{Nenkova}, Maia and {Sirocky}, Matthew M. and {Ivezi{\'c}}, {\v{Z}}eljko and {Elitzur}, Moshe},
        title = "{AGN Dusty Tori. I. Handling of Clumpy Media}",
      journal = {\apj},
         year = 2008,
        month = sep,
       volume = {685},
       number = {1},
        pages = {147-159},
          doi = {10.1086/590482},
archivePrefix = {arXiv},
       eprint = {0806.0511},
 primaryClass = {astro-ph},
       adsurl = {https://ui.adsabs.harvard.edu/abs/2008ApJ...685..147N}
}

@ARTICLE{Nenkova2008b,
       author = {{Nenkova}, Maia and {Sirocky}, Matthew M. and {Nikutta}, Robert and {Ivezi{\'c}}, {\v{Z}}eljko and {Elitzur}, Moshe},
        title = "{AGN Dusty Tori. II. Observational Implications of Clumpiness}",
      journal = {\apj},
         year = 2008,
        month = sep,
       volume = {685},
       number = {1},
        pages = {160-180},
          doi = {10.1086/590483},
archivePrefix = {arXiv},
       eprint = {0806.0512},
 primaryClass = {astro-ph},
       adsurl = {https://ui.adsabs.harvard.edu/abs/2008ApJ...685..160N}
}

@ARTICLE{Choi2016,
       author = {{Choi}, Jieun and {Dotter}, Aaron and {Conroy}, Charlie and {Cantiello}, Matteo and {Paxton}, Bill and {Johnson}, Benjamin D.},
        title = "{Mesa Isochrones and Stellar Tracks (MIST). I. Solar-scaled Models}",
      journal = {\apj},
         year = 2016,
        month = jun,
       volume = {823},
       number = {2},
          eid = {102},
        pages = {102},
          doi = {10.3847/0004-637X/823/2/102},
archivePrefix = {arXiv},
       eprint = {1604.08592},
 primaryClass = {astro-ph.SR},
       adsurl = {https://ui.adsabs.harvard.edu/abs/2016ApJ...823..102C}
}

@ARTICLE{Dotter2016,
       author = {{Dotter}, Aaron},
        title = "{MESA Isochrones and Stellar Tracks (MIST) 0: Methods for the Construction of Stellar Isochrones}",
      journal = {\apjs},
         year = 2016,
        month = jan,
       volume = {222},
       number = {1},
          eid = {8},
        pages = {8},
          doi = {10.3847/0067-0049/222/1/8},
archivePrefix = {arXiv},
       eprint = {1601.05144},
 primaryClass = {astro-ph.SR},
       adsurl = {https://ui.adsabs.harvard.edu/abs/2016ApJS..222....8D}
}

@ARTICLE{Chabrier2003,
       author = {{Chabrier}, Gilles},
        title = "{Galactic Stellar and Substellar Initial Mass Function}",
      journal = {\pasp},
         year = 2003,
        month = jul,
       volume = {115},
       number = {809},
        pages = {763-795},
          doi = {10.1086/376392},
archivePrefix = {arXiv},
       eprint = {astro-ph/0304382},
 primaryClass = {astro-ph},
       adsurl = {https://ui.adsabs.harvard.edu/abs/2003PASP..115..763C}
}

@ARTICLE{Carnall2019,
       author = {{Carnall}, Adam C. and {Leja}, Joel and {Johnson}, Benjamin D. and {McLure}, Ross J. and {Dunlop}, James S. and {Conroy}, Charlie},
        title = "{How to Measure Galaxy Star Formation Histories. I. Parametric Models}",
      journal = {\apj},
         year = 2019,
        month = mar,
       volume = {873},
       number = {1},
          eid = {44},
        pages = {44},
          doi = {10.3847/1538-4357/ab04a2},
archivePrefix = {arXiv},
       eprint = {1811.03635},
 primaryClass = {astro-ph.GA},
       adsurl = {https://ui.adsabs.harvard.edu/abs/2019ApJ...873...44C}
}

@ARTICLE{Leja2019,
       author = {{Leja}, Joel and {Carnall}, Adam C. and {Johnson}, Benjamin D. and {Conroy}, Charlie and {Speagle}, Joshua S.},
        title = "{How to Measure Galaxy Star Formation Histories. II. Nonparametric Models}",
      journal = {\apj},
         year = 2019,
        month = may,
       volume = {876},
       number = {1},
          eid = {3},
        pages = {3},
          doi = {10.3847/1538-4357/ab133c},
archivePrefix = {arXiv},
       eprint = {1811.03637},
 primaryClass = {astro-ph.GA},
       adsurl = {https://ui.adsabs.harvard.edu/abs/2019ApJ...876....3L}
}

@ARTICLE{Lower2020,
       author = {{Lower}, Sidney and {Narayanan}, Desika and {Leja}, Joel and {Johnson}, Benjamin D. and {Conroy}, Charlie and {Dav{\'e}}, Romeel},
        title = "{How Well Can We Measure the Stellar Mass of a Galaxy: The Impact of the Assumed Star Formation History Model in SED Fitting}",
      journal = {\apj},
         year = 2020,
        month = nov,
       volume = {904},
       number = {1},
          eid = {33},
        pages = {33},
          doi = {10.3847/1538-4357/abbfa7},
archivePrefix = {arXiv},
       eprint = {2006.03599},
 primaryClass = {astro-ph.GA},
       adsurl = {https://ui.adsabs.harvard.edu/abs/2020ApJ...904...33L}
}

@ARTICLE{Cannarozzo2020,
       author = {{Cannarozzo}, Carlo and {Sonnenfeld}, Alessandro and {Nipoti}, Carlo},
        title = "{The cosmic evolution of the stellar mass-velocity dispersion relation of early-type galaxies}",
      journal = {\mnras},
         year = 2020,
        month = oct,
       volume = {498},
       number = {1},
        pages = {1101-1120},
          doi = {10.1093/mnras/staa2147},
archivePrefix = {arXiv},
       eprint = {1910.06987},
 primaryClass = {astro-ph.GA},
       adsurl = {https://ui.adsabs.harvard.edu/abs/2020MNRAS.498.1101C}
}

@ARTICLE{Draine2007,
       author = {{Draine}, B.~T. and {Li}, Aigen},
        title = "{Infrared Emission from Interstellar Dust. IV. The Silicate-Graphite-PAH Model in the Post-Spitzer Era}",
      journal = {\apj},
         year = 2007,
        month = mar,
       volume = {657},
       number = {2},
        pages = {810-837},
          doi = {10.1086/511055},
archivePrefix = {arXiv},
       eprint = {astro-ph/0608003},
 primaryClass = {astro-ph},
       adsurl = {https://ui.adsabs.harvard.edu/abs/2007ApJ...657..810D}
}

@ARTICLE{Richardson2014,
       author = {{Richardson}, Chris T. and {Allen}, James T. and {Baldwin}, Jack A. and {Hewett}, Paul C. and {Ferland}, Gary J.},
        title = "{Interpreting the ionization sequence in AGN emission-line spectra}",
      journal = {\mnras},
         year = 2014,
        month = jan,
       volume = {437},
       number = {3},
        pages = {2376-2403},
          doi = {10.1093/mnras/stt2056},
archivePrefix = {arXiv},
       eprint = {1310.6402},
 primaryClass = {astro-ph.CO},
       adsurl = {https://ui.adsabs.harvard.edu/abs/2014MNRAS.437.2376R}
}

@ARTICLE{CharlotFall2000,
       author = {{Charlot}, St{\'e}phane and {Fall}, S. Michael},
        title = "{A Simple Model for the Absorption of Starlight by Dust in Galaxies}",
      journal = {\apj},
         year = 2000,
        month = aug,
       volume = {539},
       number = {2},
        pages = {718-731},
          doi = {10.1086/309250},
archivePrefix = {arXiv},
       eprint = {astro-ph/0003128},
 primaryClass = {astro-ph},
       adsurl = {https://ui.adsabs.harvard.edu/abs/2000ApJ...539..718C}
}

@ARTICLE{Speagle2020,
       author = {{Speagle}, Joshua S.},
        title = "{DYNESTY: a dynamic nested sampling package for estimating Bayesian posteriors and evidences}",
      journal = {\mnras},
         year = 2020,
        month = apr,
       volume = {493},
       number = {3},
        pages = {3132-3158},
          doi = {10.1093/mnras/staa278},
archivePrefix = {arXiv},
       eprint = {1904.02180},
 primaryClass = {astro-ph.IM},
       adsurl = {https://ui.adsabs.harvard.edu/abs/2020MNRAS.493.3132S}
}

@ARTICLE{Tanaka2026,
       author = {{Tanaka}, Takumi S. and {Silverman}, John D. and {Shimasaku}, Kazuhiro and {Jahnke}, Knud and {Li}, Junyao and {Ando}, Makoto},
        title = "{Merger-driven buildup of the M$_{BH}$─M$_{*}$ relation bridging high-z overmassive black holes with the local relation}",
      journal = {\pasj},
         year = 2026,
        month = jun,
       volume = {78},
       number = {3},
        pages = {1084-1092},
          doi = {10.1093/pasj/psag044},
archivePrefix = {arXiv},
       eprint = {2603.11747},
 primaryClass = {astro-ph.GA},
       adsurl = {https://ui.adsabs.harvard.edu/abs/2026PASJ...78.1084T}
}

@ARTICLE{Hviding2025,
       author = {{Hviding}, Raphael E. and {de Graaff}, Anna and {Miller}, Tim B. and {Setton}, David J. and {Greene}, Jenny E. and {Labb{\'e}}, Ivo and {Brammer}, Gabriel and {Bezanson}, Rachel and {Boogaard}, Leindert A. and {Cleri}, Nikko J. and {Leja}, Joel and {Maseda}, Michael V. and {McConachie}, Ian and {Matthee}, Jorryt and {Naidu}, Rohan P. and {Oesch}, Pascal A. and {Wang}, Bingjie and {Whitaker}, Katherine E. and {Williams}, Christina C.},
        title = "{RUBIES: A spectroscopic census of little red dots: All point sources with v-shaped continua have broad lines}",
      journal = {\aap},
         year = 2025,
        month = oct,
       volume = {702},
          eid = {A57},
        pages = {A57},
          doi = {10.1051/0004-6361/202555816},
archivePrefix = {arXiv},
       eprint = {2506.05459},
 primaryClass = {astro-ph.GA},
       adsurl = {https://ui.adsabs.harvard.edu/abs/2025A&A...702A..57H}
}

@ARTICLE{Silverman2025,
       author = {{Silverman}, John David and {Li}, Junyao and {Ding}, Xuheng and {Onoue}, Masafusa and {Strauss}, Michael A. and {Matsuoka}, Yoshiki and {Izumi}, Takuma and {Jahnke}, Knud and {Treu}, Tommaso and {Volonteri}, Marta and {Phillips}, Camryn L. and {Andika}, Irham T. and {Aoki}, Kentaro and {Arita}, Junya and {Baba}, Shunsuke and {Bosman}, Sarah E.~I. and {Eilers}, Anna-Christina and {Fan}, Xiaohui and {Fujimoto}, Seiji and {Habouzit}, Melanie and {Haiman}, Zoltan and {Imanishi}, Masatoshi and {Inayoshi}, Kohei and {Iwasawa}, Kazushi and {Kashikawa}, Nobunari and {Kawaguchi}, Toshihiro and {Lee}, Chien-Hsiu and {Lupi}, Alessandro and {Nagao}, Tohru and {Schindler}, Jan-Torge and {Schramm}, Malte and {Shimasaku}, Kazuhiro and {Toba}, Yoshiki and {Trakhtenbrot}, Benny and {Umehata}, Hideki and {Vestergaard}, Marianne and {Walter}, Fabian and {Wang}, Feige and {Yang}, Jinyi},
        title = "{SHELLQs─JWST Perspective on the Intrinsic Mass Relation between Supermassive Black Holes and Their Host Galaxies at z > 6}",
      journal = {\apjl},
         year = 2025,
        month = dec,
       volume = {995},
       number = {2},
          eid = {L67},
        pages = {L67},
          doi = {10.3847/2041-8213/ae279c},
archivePrefix = {arXiv},
       eprint = {2507.23066},
 primaryClass = {astro-ph.GA},
       adsurl = {https://ui.adsabs.harvard.edu/abs/2025ApJ...995L..67S}
}

@ARTICLE{Terrazas2016,
       author = {{Terrazas}, Bryan A. and {Bell}, Eric F. and {Henriques}, Bruno M.~B. and {White}, Simon D.~M. and {Cattaneo}, Andrea and {Woo}, Joanna},
        title = "{Quiescence Correlates Strongly with Directly Measured Black Hole Mass in Central Galaxies}",
      journal = {\apjl},
         year = 2016,
        month = oct,
       volume = {830},
       number = {1},
          eid = {L12},
        pages = {L12},
          doi = {10.3847/2041-8205/830/1/L12},
archivePrefix = {arXiv},
       eprint = {1609.07141},
 primaryClass = {astro-ph.GA},
       adsurl = {https://ui.adsabs.harvard.edu/abs/2016ApJ...830L..12T}
}

@ARTICLE{Piotrowska2022,
       author = {{Piotrowska}, Joanna M. and {Bluck}, Asa F.~L. and {Maiolino}, Roberto and {Peng}, Yingjie},
        title = "{On the quenching of star formation in observed and simulated central galaxies: evidence for the role of integrated AGN feedback}",
      journal = {\mnras},
         year = 2022,
        month = may,
       volume = {512},
       number = {1},
        pages = {1052-1090},
          doi = {10.1093/mnras/stab3673},
archivePrefix = {arXiv},
       eprint = {2112.07672},
 primaryClass = {astro-ph.GA},
       adsurl = {https://ui.adsabs.harvard.edu/abs/2022MNRAS.512.1052P}
}

@ARTICLE{Bhowmick2024,
       author = {{Bhowmick}, Aklant K. and {Blecha}, Laura and {Torrey}, Paul and {Kelley}, Luke Zoltan and {Weinberger}, Rainer and {Vogelsberger}, Mark and {Hernquist}, Lars and {Somerville}, Rachel S. and {Evans}, Analis Eolyn},
        title = "{Introducing the BRAHMA simulation suite: signatures of low-mass black hole seeding models in cosmological simulations}",
      journal = {\mnras},
         year = 2024,
        month = jul,
       volume = {531},
       number = {4},
        pages = {4311-4335},
          doi = {10.1093/mnras/stae1386},
archivePrefix = {arXiv},
       eprint = {2402.03626},
 primaryClass = {astro-ph.GA},
       adsurl = {https://ui.adsabs.harvard.edu/abs/2024MNRAS.531.4311B}
}

@ARTICLE{Ding2023,
       author = {{Ding}, Xuheng and {Onoue}, Masafusa and {Silverman}, John D. and {Matsuoka}, Yoshiki and {Izumi}, Takuma and {Strauss}, Michael A. and {Jahnke}, Knud and {Phillips}, Camryn L. and {Li}, Junyao and {Volonteri}, Marta and {Haiman}, Zoltan and {Andika}, Irham Taufik and {Aoki}, Kentaro and {Baba}, Shunsuke and {Bieri}, Rebekka and {Bosman}, Sarah E.~I. and {Bottrell}, Connor and {Eilers}, Anna-Christina and {Fujimoto}, Seiji and {Habouzit}, Melanie and {Imanishi}, Masatoshi and {Inayoshi}, Kohei and {Iwasawa}, Kazushi and {Kashikawa}, Nobunari and {Kawaguchi}, Toshihiro and {Kohno}, Kotaro and {Lee}, Chien-Hsiu and {Lupi}, Alessandro and {Lyu}, Jianwei and {Nagao}, Tohru and {Overzier}, Roderik and {Schindler}, Jan-Torge and {Schramm}, Malte and {Shimasaku}, Kazuhiro and {Toba}, Yoshiki and {Trakhtenbrot}, Benny and {Trebitsch}, Maxime and {Treu}, Tommaso and {Umehata}, Hideki and {Venemans}, Bram P. and {Vestergaard}, Marianne and {Walter}, Fabian and {Wang}, Feige and {Yang}, Jinyi},
        title = "{Detection of stellar light from quasar host galaxies at redshifts above 6}",
      journal = {\nat},
         year = 2023,
        month = sep,
       volume = {621},
       number = {7977},
        pages = {51-55},
          doi = {10.1038/s41586-023-06345-5},
archivePrefix = {arXiv},
       eprint = {2211.14329},
 primaryClass = {astro-ph.GA},
       adsurl = {https://ui.adsabs.harvard.edu/abs/2023Natur.621...51D}
}

@ARTICLE{Onoue2024,
       author = {{Onoue}, M. and {Ding}, X. and {Silverman}, J. D. and {Matsuoka}, Y. and {Izumi}, T. and {Strauss}, M. A. and {Ward}, C. and {Phillips}, C. L. and {Ito}, K. and {Andika}, I. T. and et al.},
        title = "{A post-starburst pathway for the formation of massive galaxies and black holes at z {\ensuremath{>}} 6}",
      journal = {Nature Astronomy},
         year = 2025,
        month = oct,
       volume = {9},
       number = {10},
        pages = {1541--1552},
          doi = {10.1038/s41550-025-02628-1},
archivePrefix = {arXiv},
       eprint = {2409.07113},
 primaryClass = {astro-ph.GA}
}

@ARTICLE{Nanayakkara2024,
       author = {{Nanayakkara}, Themiya and {Glazebrook}, Karl and {Jacobs}, Colin and {Kawinwanichakij}, Lalitwadee and {Schreiber}, Corentin and {Brammer}, Gabriel and {Esdaile}, James and {Kacprzak}, Glenn G. and {Labbe}, Ivo and {Lagos}, Claudia and {Marchesini}, Danilo and {Marsan}, Z. Cemile and {Oesch}, Pascal A. and {Papovich}, Casey and {Remus}, Rhea-Silvia and {Tran}, Kim-Vy H.},
        title = "{A population of faint, old, and massive quiescent galaxies at 3 <z <4 revealed by JWST NIRSpec Spectroscopy}",
      journal = {Scientific Reports},
         year = 2024,
        month = feb,
       volume = {14},
          eid = {3724},
        pages = {3724},
          doi = {10.1038/s41598-024-52585-4},
archivePrefix = {arXiv},
       eprint = {2212.11638},
 primaryClass = {astro-ph.GA},
       adsurl = {https://ui.adsabs.harvard.edu/abs/2024NatSR..14.3724N}
}

@ARTICLE{Peng2007,
       author = {{Peng}, Chien Y.},
        title = "{How Mergers May Affect the Mass Scaling Relation between Gravitationally Bound Systems}",
      journal = {\apj},
         year = 2007,
        month = dec,
       volume = {671},
       number = {2},
        pages = {1098-1107},
          doi = {10.1086/522774},
archivePrefix = {arXiv},
       eprint = {0704.1860},
 primaryClass = {astro-ph},
       adsurl = {https://ui.adsabs.harvard.edu/abs/2007ApJ...671.1098P}
}

@ARTICLE{Fabian2012,
       author = {{Fabian}, A.~C.},
        title = "{Observational Evidence of Active Galactic Nuclei Feedback}",
      journal = {\araa},
         year = 2012,
        month = sep,
       volume = {50},
        pages = {455-489},
          doi = {10.1146/annurev-astro-081811-125521},
archivePrefix = {arXiv},
       eprint = {1204.4114},
 primaryClass = {astro-ph.CO},
       adsurl = {https://ui.adsabs.harvard.edu/abs/2012ARA&A..50..455F}
}

@ARTICLE{Jahnke2011,
       author = {{Jahnke}, Knud and {Macci{\`o}}, Andrea V.},
        title = "{The Non-causal Origin of the Black-hole-galaxy Scaling Relations}",
      journal = {\apj},
         year = 2011,
        month = jun,
       volume = {734},
       number = {2},
          eid = {92},
        pages = {92},
          doi = {10.1088/0004-637X/734/2/92},
archivePrefix = {arXiv},
       eprint = {1006.0482},
 primaryClass = {astro-ph.CO},
       adsurl = {https://ui.adsabs.harvard.edu/abs/2011ApJ...734...92J}
}

@ARTICLE{Carnall2023,
       author = {{Carnall}, A.~C. and {McLeod}, D.~J. and {McLure}, R.~J. and {Dunlop}, J.~S. and {Begley}, R. and {Cullen}, F. and {Donnan}, C.~T. and {Hamadouche}, M.~L. and {Jewell}, S.~M. and {Jones}, E.~W. and {Pollock}, C.~L. and {Wild}, V.},
        title = "{A surprising abundance of massive quiescent galaxies at 3 < z < 5 in the first data from JWST CEERS}",
      journal = {\mnras},
         year = 2023,
        month = apr,
       volume = {520},
       number = {3},
        pages = {3974-3985},
          doi = {10.1093/mnras/stad369},
archivePrefix = {arXiv},
       eprint = {2208.00986},
 primaryClass = {astro-ph.GA},
       adsurl = {https://ui.adsabs.harvard.edu/abs/2023MNRAS.520.3974C}
}

@ARTICLE{Deng2024,
       author = {{Deng}, Yunwei and {Li}, Hui and {Liu}, Boyuan and {Kannan}, Rahul and {Smith}, Aaron and {Bryan}, Greg L.},
        title = "{RIGEL: Simulating dwarf galaxies at solar mass resolution with radiative transfer and feedback from individual massive stars}",
      journal = {\aap},
         year = 2024,
        month = nov,
       volume = {691},
          eid = {A231},
        pages = {A231},
          doi = {10.1051/0004-6361/202450699},
archivePrefix = {arXiv},
       eprint = {2405.08869},
 primaryClass = {astro-ph.GA},
       adsurl = {https://ui.adsabs.harvard.edu/abs/2024A&A...691A.231D}
}

@ARTICLE{Habouzit2017,
       author = {{Habouzit}, M{\'e}lanie and {Volonteri}, Marta and {Dubois}, Yohan},
        title = "{Blossoms from black hole seeds: properties and early growth regulated by supernova feedback}",
      journal = {\mnras},
         year = 2017,
        month = jul,
       volume = {468},
       number = {4},
        pages = {3935-3948},
          doi = {10.1093/mnras/stx666},
archivePrefix = {arXiv},
       eprint = {1605.09394},
 primaryClass = {astro-ph.GA},
       adsurl = {https://ui.adsabs.harvard.edu/abs/2017MNRAS.468.3935H}
}

@ARTICLE{Baggen2023,
       author = {{Baggen}, Josephine F.~W. and {van Dokkum}, Pieter and {Labb{\'e}}, Ivo and {Brammer}, Gabriel and {Miller}, Tim B. and {Bezanson}, Rachel and {Leja}, Joel and {Wang}, Bingjie and {Whitaker}, Katherine E. and {Suess}, Katherine A. and {Nelson}, Erica J.},
        title = "{Sizes and Mass Profiles of Candidate Massive Galaxies Discovered by JWST at 7 < z < 9: Evidence for Very Early Formation of the Central 100 pc of Present-day Ellipticals}",
      journal = {\apjl},
         year = 2023,
        month = sep,
       volume = {955},
       number = {1},
          eid = {L12},
        pages = {L12},
          doi = {10.3847/2041-8213/acf5ef},
archivePrefix = {arXiv},
       eprint = {2305.17162},
 primaryClass = {astro-ph.GA},
       adsurl = {https://ui.adsabs.harvard.edu/abs/2023ApJ...955L..12B}
}

@ARTICLE{Zhu2022,
       author = {{Zhu}, Qirong and {Li}, Yuexing and {Li}, Yiting and {Maji}, Moupiya and {Yajima}, Hidenobu and {Schneider}, Raffaella and {Hernquist}, Lars},
        title = "{The formation of the first quasars: the black hole seeds, accretion, and feedback models}",
      journal = {\mnras},
         year = 2022,
        month = aug,
       volume = {514},
       number = {4},
        pages = {5583-5606},
          doi = {10.1093/mnras/stac1556},
archivePrefix = {arXiv},
       eprint = {2012.01458},
 primaryClass = {astro-ph.GA},
       adsurl = {https://ui.adsabs.harvard.edu/abs/2022MNRAS.514.5583Z}
}

@ARTICLE{Baker2024,
       author = {{Baker}, W. M. and {Lim}, S. and {D'Eugenio}, F. and {Maiolino}, R. and {Ji}, Z. and {Arribas}, S. and {Bunker}, A. J. and {Carniani}, S. and {Charlot}, S. and {de Graaff}, A. and et al.},
        title = "{The abundance and nature of high-redshift quiescent galaxies from JADES spectroscopy and the FLAMINGO simulations}",
      journal = {\mnras},
         year = 2025,
        month = apr,
       volume = {539},
       number = {1},
        pages = {557--589},
          doi = {10.1093/mnras/staf475},
archivePrefix = {arXiv},
       eprint = {2410.14773},
 primaryClass = {astro-ph.GA}
}

@ARTICLE{Volonteri2023,
       author = {{Volonteri}, Marta and {Habouzit}, M{\'e}lanie and {Colpi}, Monica},
        title = "{What if young z > 9 JWST galaxies hosted massive black holes?}",
      journal = {\mnras},
         year = 2023,
        month = may,
       volume = {521},
       number = {1},
        pages = {241-250},
          doi = {10.1093/mnras/stad499},
archivePrefix = {arXiv},
       eprint = {2212.04710},
 primaryClass = {astro-ph.GA},
       adsurl = {https://ui.adsabs.harvard.edu/abs/2023MNRAS.521..241V}
}

@ARTICLE{Volonteri2010,
       author = {{Volonteri}, Marta},
        title = "{Formation of supermassive black holes}",
      journal = {\aapr},
         year = 2010,
        month = jul,
       volume = {18},
       number = {3},
        pages = {279-315},
          doi = {10.1007/s00159-010-0029-x},
archivePrefix = {arXiv},
       eprint = {1003.4404},
 primaryClass = {astro-ph.CO},
       adsurl = {https://ui.adsabs.harvard.edu/abs/2010A&ARv..18..279V}
}

@ARTICLE{Maseda2024,
       author = {{Maseda}, Michael V. and {de Graaff}, Anna and {Franx}, Marijn and {Rix}, Hans-Walter and {Carniani}, Stefano and {Laseter}, Isaac and {Dudzevi{\v{c}}i{\={u}}t{\.{e}}}, Ugn{\.{e}} and {Rawle}, Tim and {Parlanti}, Eleonora and {Arribas}, Santiago and {Bunker}, Andrew J. and {Cameron}, Alex J. and {Charlot}, Stephane and {Curti}, Mirko and {D'Eugenio}, Francesco and {Jones}, Gareth C. and {Kumari}, Nimisha and {Maiolino}, Roberto and {{\"U}bler}, Hannah and {Saxena}, Aayush and {Smit}, Renske and {Willott}, Chris and {Witstok}, Joris},
        title = "{The NIRSpec Wide GTO Survey}",
      journal = {\aap},
         year = 2024,
        month = sep,
       volume = {689},
          eid = {A73},
        pages = {A73},
          doi = {10.1051/0004-6361/202449914},
archivePrefix = {arXiv},
       eprint = {2403.05506},
 primaryClass = {astro-ph.GA},
       adsurl = {https://ui.adsabs.harvard.edu/abs/2024A&A...689A..73M}
}

@ARTICLE{Eisenstein2023,
       author = {{Eisenstein}, D. J. and {Willott}, C. and {Alberts}, S. and {Arribas}, S. and {Bonaventura}, N. and {Bunker}, A. J. and {Cameron}, A. J. and {Carniani}, S. and {Charlot}, S. and {Curtis-Lake}, E. and et al.},
        title = "{Overview of the JWST Advanced Deep Extragalactic Survey (JADES)}",
      journal = {\apjs},
         year = 2026,
        month = mar,
       volume = {283},
       number = {1},
        pages = {6},
          doi = {10.3847/1538-4365/ae3163},
archivePrefix = {arXiv},
       eprint = {2306.02465},
 primaryClass = {astro-ph.GA}
}

@ARTICLE{Li2021mass,
       author = {{Li}, Junyao and {Silverman}, John D. and {Ding}, Xuheng and {Strauss}, Michael A. and {Goulding}, Andy and {Schramm}, Malte and {Yesuf}, Hassen M. and {Sun}, Mouyuan and {Xue}, Yongquan and {Birrer}, Simon and {Shi}, Jingjing and {Toba}, Yoshiki and {Nagao}, Tohru and {Imanishi}, Masatoshi},
        title = "{Synchronized Coevolution between Supermassive Black Holes and Galaxies over the Last Seven Billion Years as Revealed by Hyper Suprime-Cam}",
      journal = {\apj},
         year = 2021,
        month = dec,
       volume = {922},
       number = {2},
          eid = {142},
        pages = {142},
          doi = {10.3847/1538-4357/ac2301},
archivePrefix = {arXiv},
       eprint = {2109.02751},
 primaryClass = {astro-ph.GA},
       adsurl = {https://ui.adsabs.harvard.edu/abs/2021ApJ...922..142L}
}

@ARTICLE{Fan2023,
       author = {{Fan}, Xiaohui and {Ba{\~n}ados}, Eduardo and {Simcoe}, Robert A.},
        title = "{Quasars and the Intergalactic Medium at Cosmic Dawn}",
      journal = {\araa},
         year = 2023,
        month = aug,
       volume = {61},
        pages = {373-426},
          doi = {10.1146/annurev-astro-052920-102455},
archivePrefix = {arXiv},
       eprint = {2212.06907},
 primaryClass = {astro-ph.GA},
       adsurl = {https://ui.adsabs.harvard.edu/abs/2023ARA&A..61..373F}
}

@ARTICLE{Lauer2007,
       author = {{Lauer}, Tod R. and {Tremaine}, Scott and {Richstone}, Douglas and {Faber}, S.~M.},
        title = "{Selection Bias in Observing the Cosmological Evolution of the M$_{{\textbullet}}$-{\ensuremath{\sigma}} and M$_{{\textbullet}}$-L Relationships}",
      journal = {\apj},
         year = 2007,
        month = nov,
       volume = {670},
       number = {1},
        pages = {249-260},
          doi = {10.1086/522083},
archivePrefix = {arXiv},
       eprint = {0705.4103},
 primaryClass = {astro-ph},
       adsurl = {https://ui.adsabs.harvard.edu/abs/2007ApJ...670..249L}
}

@ARTICLE{Boyett2024,
       author = {{Boyett}, Kit and {Bunker}, Andrew J. and {Curtis-Lake}, Emma and {Chevallard}, Jacopo and {Cameron}, Alex J. and {Jones}, Gareth C. and {Saxena}, Aayush and {Charlot}, St{\'e}phane and {Curti}, Mirko and {Wallace}, Imaan E.~B. and {Arribas}, Santiago and {Carniani}, Stefano and {Willott}, Chris and {Alberts}, Stacey and {Eisenstein}, Daniel J. and {Hainline}, Kevin and {Hausen}, Ryan and {Johnson}, Benjamin D. and {Rieke}, Marcia and {Robertson}, Brant and {Stark}, Daniel P. and {Tacchella}, Sandro and {Williams}, Christina C. and {Chen}, Zuyi and {Egami}, Eiichi and {Endsley}, Ryan and {Kumari}, Nimisha and {Laseter}, Isaac and {Looser}, Tobias J. and {Maseda}, Michael V. and {Scholtz}, Jan and {Shivaei}, Irene and {Simmonds}, Charlotte and {Smit}, Renske and {{\"U}bler}, Hannah and {Witstok}, Joris},
        title = "{Extreme emission line galaxies detected in JADES JWST/NIRSpec - I. Inferred galaxy properties}",
      journal = {\mnras},
         year = 2024,
        month = dec,
       volume = {535},
       number = {2},
        pages = {1796-1828},
          doi = {10.1093/mnras/stae2430},
archivePrefix = {arXiv},
       eprint = {2401.16934},
 primaryClass = {astro-ph.GA},
       adsurl = {https://ui.adsabs.harvard.edu/abs/2024MNRAS.535.1796B}
}

@ARTICLE{deGraaff2024qs,
       author = {{de Graaff}, A. and {Setton}, D. J. and {Brammer}, G. and {Cutler}, S. and {Suess}, K. A. and {Labb{\'e}}, I. and {Leja}, J. and {Weibel}, A. and {Maseda}, M. V. and {Whitaker}, K. E. and et al.},
        title = "{Efficient formation of a massive quiescent galaxy at redshift 4.9}",
      journal = {Nature Astronomy},
         year = 2025,
        month = feb,
       volume = {9},
       number = {2},
        pages = {280--292},
          doi = {10.1038/s41550-024-02424-3},
archivePrefix = {arXiv},
       eprint = {2404.05683},
 primaryClass = {astro-ph.GA}
}

@ARTICLE{Arteaga2024,
       author = {{Gim{\'e}nez-Arteaga}, C. and {Fujimoto}, S. and {Valentino}, F. and {Brammer}, G. B. and {Mason}, C. A. and {Rizzo}, F. and {Rusakov}, V. and {Colina}, L. and {Prieto-Lyon}, G. and {Oesch}, P. A. and et al.},
        title = "{Outshining in the spatially resolved analysis of a strongly lensed galaxy at z = 6.072 with JWST NIRCam}",
      journal = {\aap},
         year = 2024,
        month = jun,
       volume = {686},
        pages = {A63},
          doi = {10.1051/0004-6361/202349135},
archivePrefix = {arXiv},
       eprint = {2402.17875},
 primaryClass = {astro-ph.GA}
}

@ARTICLE{Gimenez2023,
       author = {{Gim{\'e}nez-Arteaga}, Clara and {Oesch}, Pascal A. and {Brammer}, Gabriel B. and {Valentino}, Francesco and {Mason}, Charlotte A. and {Weibel}, Andrea and {Barrufet}, Laia and {Fujimoto}, Seiji and {Heintz}, Kasper E. and {Nelson}, Erica J. and {Strait}, Victoria B. and {Suess}, Katherine A. and {Gibson}, Justus},
        title = "{Spatially Resolved Properties of Galaxies at 5 < z < 9 in the SMACS 0723 JWST ERO Field}",
      journal = {\apj},
         year = 2023,
        month = may,
       volume = {948},
       number = {2},
          eid = {126},
        pages = {126},
          doi = {10.3847/1538-4357/acc5ea},
archivePrefix = {arXiv},
       eprint = {2212.08670},
 primaryClass = {astro-ph.GA},
       adsurl = {https://ui.adsabs.harvard.edu/abs/2023ApJ...948..126G}
}

@article{Raftery1995,
 ISSN = {00811750, 14679531},
 URL = {http://www.jstor.org/stable/271063},
 author = {Adrian E. Raftery},
 journal = {Sociological Methodology},
 pages = {111--163},
 publisher = {[American Sociological Association, Wiley, Sage Publications, Inc.]},
 title = {Bayesian Model Selection in Social Research},
 urldate = {2023-07-10},
 volume = {25},
 year = {1995}
}

@ARTICLE{Vestergaard2006,
       author = {{Vestergaard}, Marianne and {Peterson}, Bradley M.},
        title = "{Determining Central Black Hole Masses in Distant Active Galaxies and Quasars. II. Improved Optical and UV Scaling Relationships}",
      journal = {\apj},
         year = 2006,
        month = apr,
       volume = {641},
       number = {2},
        pages = {689-709},
          doi = {10.1086/500572},
archivePrefix = {arXiv},
       eprint = {astro-ph/0601303},
 primaryClass = {astro-ph},
       adsurl = {https://ui.adsabs.harvard.edu/abs/2006ApJ...641..689V}
}

@ARTICLE{Shen2024RM,
       author = {{Shen}, Yue and {Grier}, Catherine J. and {Horne}, Keith and {Stone}, Zachary and {Li}, Jennifer I. and {Yang}, Qian and {Homayouni}, Yasaman and {Trump}, Jonathan R. and {Anderson}, Scott F. and {Brandt}, W.~N. and {Hall}, Patrick B. and {Ho}, Luis C. and {Jiang}, Linhua and {Petitjean}, Patrick and {Schneider}, Donald P. and {Tao}, Charling and {Donnan}, Fergus. R. and {AlSayyad}, Yusra and {Bershady}, Matthew A. and {Blanton}, Michael R. and {Bizyaev}, Dmitry and {Bundy}, Kevin and {Chen}, Yuguang and {Davis}, Megan C. and {Dawson}, Kyle and {Fan}, Xiaohui and {Greene}, Jenny E. and {Gr{\"o}ller}, Hannes and {Guo}, Yucheng and {Ibarra-Medel}, H{\'e}ctor and {Jiang}, Yuanzhe and {Keenan}, Ryan P. and {Kollmeier}, Juna A. and {Lejoly}, Cassandra and {Li}, Zefeng and {de la Macorra}, Axel and {Moe}, Maxwell and {Nie}, Jundan and {Rossi}, Graziano and {Smith}, Paul S. and {Tee}, Wei Leong and {Weijmans}, Anne-Marie and {Xu}, Jiachuan and {Yue}, Minghao and {Zhou}, Xu and {Zhou}, Zhimin and {Zou}, Hu},
        title = "{The Sloan Digital Sky Survey Reverberation Mapping Project: Key Results}",
      journal = {\apjs},
         year = 2024,
        month = jun,
       volume = {272},
       number = {2},
          eid = {26},
        pages = {26},
          doi = {10.3847/1538-4365/ad3936},
archivePrefix = {arXiv},
       eprint = {2305.01014},
 primaryClass = {astro-ph.GA},
       adsurl = {https://ui.adsabs.harvard.edu/abs/2024ApJS..272...26S}
}

@MISC{lmfit,
       author = {{Newville}, Matt and {Otten}, Renee and {Nelson}, Andrew and {Stensitzki}, Till and {Ingargiola}, Antonino and {Allan}, Dan and {Fox}, Austin and {Carter}, Faustin and {Micha{\l}} and {Osborn}, Ray and {Pustakhod}, Dima and {Weigand}, Sebastian and {lneuhaus} and {Aristov}, Andrey and {Glenn} and {Mark} and {mgunyho} and {Deil}, Christoph and {Hansen}, Allan L.~R. and {Pasquevich}, Gustavo and {Foks}, Leon and {Zobrist}, Nicholas and {Frost}, Oliver and {Stuermer} and {Jaskula}, Jean-Christophe and {Caldwell}, Shane and {Eendebak}, Pieter and {Pompili}, Matteo and {Hedegaard Nielsen}, Jens and {Persaud}, Arun},
        title = "{lmfit/lmfit-py: 1.3.2}",
         year = 2024,
        month = jul,
          eid = {10.5281/zenodo.12785036},
          doi = {10.5281/zenodo.12785036},
      version = {1.3.2},
    publisher = {Zenodo},
 howpublished = {Zenodo},
       adsurl = {https://ui.adsabs.harvard.edu/abs/2024zndo..12785036N}
}

@ARTICLE{Calzetti2000,
       author = {{Calzetti}, Daniela and {Armus}, Lee and {Bohlin}, Ralph C. and {Kinney}, Anne L. and {Koornneef}, Jan and {Storchi-Bergmann}, Thaisa},
        title = "{The Dust Content and Opacity of Actively Star-forming Galaxies}",
      journal = {\apj},
         year = 2000,
        month = apr,
       volume = {533},
       number = {2},
        pages = {682-695},
          doi = {10.1086/308692},
archivePrefix = {arXiv},
       eprint = {astro-ph/9911459},
 primaryClass = {astro-ph},
       adsurl = {https://ui.adsabs.harvard.edu/abs/2000ApJ...533..682C}
}

@ARTICLE{Prevot1984,
       author = {{Prevot}, M.~L. and {Lequeux}, J. and {Maurice}, E. and {Prevot}, L. and {Rocca-Volmerange}, B.},
        title = "{The typical interstellar extinction in the Small Magellanic Cloud.}",
      journal = {\aap},
         year = 1984,
        month = mar,
       volume = {132},
        pages = {389-392},
       adsurl = {https://ui.adsabs.harvard.edu/abs/1984A&A...132..389P}
}

@ARTICLE{Horne1986,
       author = {{Horne}, K.},
        title = "{An optimal extraction algorithm for CCD spectroscopy.}",
      journal = {\pasp},
         year = 1986,
        month = jun,
       volume = {98},
        pages = {609-617},
          doi = {10.1086/131801},
       adsurl = {https://ui.adsabs.harvard.edu/abs/1986PASP...98..609H}
}

@MISC{msaexp,
       author = {{Brammer}, Gabriel},
        title = "{msaexp: NIRSpec analyis tools}",
         year = 2023,
        month = sep,
          eid = {10.5281/zenodo.7299500},
          doi = {10.5281/zenodo.7299500},
      version = {0.6.17},
    publisher = {Zenodo},
 howpublished = {Zenodo},
       adsurl = {https://ui.adsabs.harvard.edu/abs/2022zndo...7299500B}
}

@ARTICLE{Wang2024LRD,
       author = {{Wang}, Bingjie and {Leja}, Joel and {de Graaff}, Anna and {Brammer}, Gabriel B. and {Weibel}, Andrea and {van Dokkum}, Pieter and {Baggen}, Josephine F.~W. and {Suess}, Katherine A. and {Greene}, Jenny E. and {Bezanson}, Rachel and {Cleri}, Nikko J. and {Hirschmann}, Michaela and {Labb{\'e}}, Ivo and {Matthee}, Jorryt and {McConachie}, Ian and {Naidu}, Rohan P. and {Nelson}, Erica and {Oesch}, Pascal A. and {Setton}, David J. and {Williams}, Christina C.},
        title = "{RUBIES: Evolved Stellar Populations with Extended Formation Histories at z {\ensuremath{\sim}} 7{\textendash}8 in Candidate Massive Galaxies Identified with JWST/NIRSpec}",
      journal = {\apjl},
         year = 2024,
        month = jul,
       volume = {969},
       number = {1},
          eid = {L13},
        pages = {L13},
          doi = {10.3847/2041-8213/ad55f7},
archivePrefix = {arXiv},
       eprint = {2405.01473},
 primaryClass = {astro-ph.GA},
       adsurl = {https://ui.adsabs.harvard.edu/abs/2024ApJ...969L..13W}
}

@ARTICLE{Taylor2024,
       author = {{Taylor}, A. J. and {Finkelstein}, S. L. and {Kocevski}, D. D. and {Jeon}, J. and {Bromm}, V. and {Amor{\'{\i}}n}, R. O. and {Arrabal Haro}, P. and {Backhaus}, B. E. and {Bagley}, M. B. and {Banados}, E. and et al.},
        title = "{Broad-line AGNs at 3.5 {\ensuremath{<}} z {\ensuremath{<}} 6: The Black Hole Mass Function and a Connection with Little Red Dots}",
      journal = {\apj},
         year = 2025,
        month = jun,
       volume = {986},
       number = {2},
        pages = {165},
          doi = {10.3847/1538-4357/add15b},
archivePrefix = {arXiv},
       eprint = {2409.06772},
 primaryClass = {astro-ph.GA}
}

@ARTICLE{Richards2006,
       author = {{Richards}, Gordon T. and {Lacy}, Mark and {Storrie-Lombardi}, Lisa J. and {Hall}, Patrick B. and {Gallagher}, S.~C. and {Hines}, Dean C. and {Fan}, Xiaohui and {Papovich}, Casey and {Vanden Berk}, Daniel E. and {Trammell}, George B. and {Schneider}, Donald P. and {Vestergaard}, Marianne and {York}, Donald G. and {Jester}, Sebastian and {Anderson}, Scott F. and {Budav{\'a}ri}, Tam{\'a}s and {Szalay}, Alexander S.},
        title = "{Spectral Energy Distributions and Multiwavelength Selection of Type 1 Quasars}",
      journal = {\apjs},
         year = 2006,
        month = oct,
       volume = {166},
       number = {2},
        pages = {470-497},
          doi = {10.1086/506525},
archivePrefix = {arXiv},
       eprint = {astro-ph/0601558},
 primaryClass = {astro-ph},
       adsurl = {https://ui.adsabs.harvard.edu/abs/2006ApJS..166..470R}
}

@ARTICLE{Ito2022,
       author = {{Ito}, Kei and {Tanaka}, Masayuki and {Miyaji}, Takamitsu and {Ilbert}, Olivier and {Kauffmann}, Olivier B. and {Koekemoer}, Anton M. and {Marchesi}, Stefano and {Shuntov}, Marko and {Toft}, Sune and {Valentino}, Francesco and {Weaver}, John R.},
        title = "{COSMOS2020: Ubiquitous AGN Activity of Massive Quiescent Galaxies at 0 < z < 5 Revealed by X-Ray and Radio Stacking}",
      journal = {\apj},
         year = 2022,
        month = apr,
       volume = {929},
       number = {1},
          eid = {53},
        pages = {53},
          doi = {10.3847/1538-4357/ac5aaf},
archivePrefix = {arXiv},
       eprint = {2203.04322},
 primaryClass = {astro-ph.GA},
       adsurl = {https://ui.adsabs.harvard.edu/abs/2022ApJ...929...53I}
}

@ARTICLE{Park2024,
       author = {{Park}, Minjung and {Belli}, Sirio and {Conroy}, Charlie and {Johnson}, Benjamin D. and {Davies}, Rebecca L. and {Leja}, Joel and {Tacchella}, Sandro and {Mendel}, J. Trevor and {Benton}, Chlo{\"e} and {Bugiani}, Letizia and {Emami}, Razieh and {Khoram}, Amir H. and {Li}, Yijia and {Maheson}, Gabriel and {Mathews}, Elijah P. and {Naidu}, Rohan P. and {Nelson}, Erica J. and {Terrazas}, Bryan A. and {Weinberger}, Rainer},
        title = "{Widespread Rapid Quenching at Cosmic Noon Revealed by JWST Deep Spectroscopy}",
      journal = {\apj},
         year = 2024,
        month = nov,
       volume = {976},
       number = {1},
          eid = {72},
        pages = {72},
          doi = {10.3847/1538-4357/ad7e15},
archivePrefix = {arXiv},
       eprint = {2404.17945},
 primaryClass = {astro-ph.GA},
       adsurl = {https://ui.adsabs.harvard.edu/abs/2024ApJ...976...72P}
}

@ARTICLE{DEugenio2023,
       author = {{D'Eugenio}, F. and {P{\'e}rez-Gonz{\'a}lez}, P. G. and {Maiolino}, R. and {Scholtz}, J. and {Perna}, M. and {Circosta}, C. and {{\"U}bler}, H. and {Arribas}, S. and {B{\"o}ker}, T. and {Bunker}, A. J. and et al.},
        title = "{A fast-rotator post-starburst galaxy quenched by supermassive black-hole feedback at z = 3}",
      journal = {Nature Astronomy},
         year = 2024,
        month = nov,
       volume = {8},
       number = {11},
        pages = {1443--1456},
          doi = {10.1038/s41550-024-02345-1},
archivePrefix = {arXiv},
       eprint = {2308.06317},
 primaryClass = {astro-ph.GA}
}

@ARTICLE{Reines2015,
       author = {{Reines}, Amy E. and {Volonteri}, Marta},
        title = "{Relations between Central Black Hole Mass and Total Galaxy Stellar Mass in the Local Universe}",
      journal = {\apj},
         year = 2015,
        month = nov,
       volume = {813},
       number = {2},
          eid = {82},
        pages = {82},
          doi = {10.1088/0004-637X/813/2/82},
archivePrefix = {arXiv},
       eprint = {1508.06274},
 primaryClass = {astro-ph.GA},
       adsurl = {https://ui.adsabs.harvard.edu/abs/2015ApJ...813...82R}
}

@ARTICLE{Schulze2014,
       author = {{Schulze}, Andreas and {Wisotzki}, Lutz},
        title = "{Accounting for selection effects in the BH-bulge relations: no evidence for cosmological evolution}",
      journal = {\mnras},
         year = 2014,
        month = mar,
       volume = {438},
       number = {4},
        pages = {3422-3433},
          doi = {10.1093/mnras/stt2457},
archivePrefix = {arXiv},
       eprint = {1312.5610},
 primaryClass = {astro-ph.CO},
       adsurl = {https://ui.adsabs.harvard.edu/abs/2014MNRAS.438.3422S}
}

@ARTICLE{Li2024,
       author = {{Li}, J. and {Silverman}, J. D. and {Shen}, Y. and {Volonteri}, M. and {Jahnke}, K. and {Zhuang}, M.-Y. and {Scoggins}, M. T. and {Ding}, X. and {Harikane}, Y. and {Onoue}, M. and et al.},
        title = "{Tip of the Iceberg: Overmassive Black Holes at 4 {\ensuremath{<}} z {\ensuremath{<}} 7 Found by JWST Are Not Inconsistent with the Local M$_{\rm BH}$--M$_{\star}$ Relation}",
      journal = {\apj},
         year = 2025,
        month = mar,
       volume = {981},
       number = {1},
        pages = {19},
          doi = {10.3847/1538-4357/ada603},
archivePrefix = {arXiv},
       eprint = {2403.00074},
 primaryClass = {astro-ph.GA}
}

@ARTICLE{Ding2022,
       author = {{Ding}, Xuheng and {Silverman}, John D. and {Treu}, Tommaso and {Li}, Junyao and {Bhowmick}, Aklant K. and {Menci}, Nicola and {Volonteri}, Marta and {Blecha}, Laura and {Di Matteo}, Tiziana and {Dubois}, Yohan},
        title = "{Concordance between Observations and Simulations in the Evolution of the Mass Relation between Supermassive Black Holes and Their Host Galaxies}",
      journal = {\apj},
         year = 2022,
        month = jul,
       volume = {933},
       number = {2},
          eid = {132},
        pages = {132},
          doi = {10.3847/1538-4357/ac714c},
archivePrefix = {arXiv},
       eprint = {2205.04481},
 primaryClass = {astro-ph.GA},
       adsurl = {https://ui.adsabs.harvard.edu/abs/2022ApJ...933..132D}
}

@ARTICLE{King2015,
       author = {{King}, Andrew and {Pounds}, Ken},
        title = "{Powerful Outflows and Feedback from Active Galactic Nuclei}",
      journal = {\araa},
         year = 2015,
        month = aug,
       volume = {53},
        pages = {115-154},
          doi = {10.1146/annurev-astro-082214-122316},
archivePrefix = {arXiv},
       eprint = {1503.05206},
 primaryClass = {astro-ph.GA},
       adsurl = {https://ui.adsabs.harvard.edu/abs/2015ARA&A..53..115K}
}

@ARTICLE{Pacucci2024,
       author = {{Pacucci}, Fabio and {Loeb}, Abraham},
        title = "{The Redshift Evolution of the M $_{{\textbullet}}${\textendash}M $_{{\ensuremath{\star}}}$ Relation for JWST's Supermassive Black Holes at z > 4}",
      journal = {\apj},
         year = 2024,
        month = apr,
       volume = {964},
       number = {2},
          eid = {154},
        pages = {154},
          doi = {10.3847/1538-4357/ad3044},
archivePrefix = {arXiv},
       eprint = {2401.04159},
 primaryClass = {astro-ph.GA},
       adsurl = {https://ui.adsabs.harvard.edu/abs/2024ApJ...964..154P}
}

@ARTICLE{Dattathri2024,
       author = {{Dattathri}, S. and {Natarajan}, P. and {Porras-Valverde}, A. J. and {Burke}, C. J. and {Chen}, N. and {Di Matteo}, T. and {Ni}, Y.},
        title = "{The Redshift Evolution of the M$_{\rm BH}$--M$_{*}$ Scaling Relation: New Insights from Cosmological Simulations and Semianalytic Models}",
      journal = {\apj},
         year = 2025,
        month = may,
       volume = {984},
       number = {2},
        pages = {122},
          doi = {10.3847/1538-4357/adbeef},
archivePrefix = {arXiv},
       eprint = {2410.13958},
 primaryClass = {astro-ph.GA}
}

@ARTICLE{Matthee2024,
       author = {{Matthee}, Jorryt and {Naidu}, Rohan P. and {Brammer}, Gabriel and {Chisholm}, John and {Eilers}, Anna-Christina and {Goulding}, Andy and {Greene}, Jenny and {Kashino}, Daichi and {Labbe}, Ivo and {Lilly}, Simon J. and {Mackenzie}, Ruari and {Oesch}, Pascal A. and {Weibel}, Andrea and {Wuyts}, Stijn and {Xiao}, Mengyuan and {Bordoloi}, Rongmon and {Bouwens}, Rychard and {van Dokkum}, Pieter and {Illingworth}, Garth and {Kramarenko}, Ivan and {Maseda}, Michael V. and {Mason}, Charlotte and {Meyer}, Romain A. and {Nelson}, Erica J. and {Reddy}, Naveen A. and {Shivaei}, Irene and {Simcoe}, Robert A. and {Yue}, Minghao},
        title = "{Little Red Dots: An Abundant Population of Faint Active Galactic Nuclei at z {\ensuremath{\sim}} 5 Revealed by the EIGER and FRESCO JWST Surveys}",
      journal = {\apj},
         year = 2024,
        month = mar,
       volume = {963},
       number = {2},
          eid = {129},
        pages = {129},
          doi = {10.3847/1538-4357/ad2345},
archivePrefix = {arXiv},
       eprint = {2306.05448},
 primaryClass = {astro-ph.GA},
       adsurl = {https://ui.adsabs.harvard.edu/abs/2024ApJ...963..129M}
}

@ARTICLE{Natarajan2024,
       author = {{Natarajan}, Priyamvada and {Pacucci}, Fabio and {Ricarte}, Angelo and {Bogd{\'a}n}, {\'A}kos and {Goulding}, Andy D. and {Cappelluti}, Nico},
        title = "{First Detection of an Overmassive Black Hole Galaxy UHZ1: Evidence for Heavy Black Hole Seed Formation from Direct Collapse}",
      journal = {\apjl},
         year = 2024,
        month = jan,
       volume = {960},
       number = {1},
          eid = {L1},
        pages = {L1},
          doi = {10.3847/2041-8213/ad0e76},
archivePrefix = {arXiv},
       eprint = {2308.02654},
 primaryClass = {astro-ph.HE},
       adsurl = {https://ui.adsabs.harvard.edu/abs/2024ApJ...960L...1N}
}

@ARTICLE{Carnall2024,
       author = {{Carnall}, A.~C. and {Cullen}, F. and {McLure}, R.~J. and {McLeod}, D.~J. and {Begley}, R. and {Donnan}, C.~T. and {Dunlop}, J.~S. and {Shapley}, A.~E. and {Rowlands}, K. and {Almaini}, O. and {Arellano-C{\'o}rdova}, K.~Z. and {Barrufet}, L. and {Cimatti}, A. and {Ellis}, R.~S. and {Grogin}, N.~A. and {Hamadouche}, M.~L. and {Illingworth}, G.~D. and {Koekemoer}, A.~M. and {Leung}, H. -H. and {Lovell}, C.~C. and {P{\'e}rez-Gonz{\'a}lez}, P.~G. and {Santini}, P. and {Stanton}, T.~M. and {Wild}, V.},
        title = "{The JWST EXCELS survey: too much, too young, too fast? Ultra-massive quiescent galaxies at 3 < z < 5}",
      journal = {\mnras},
         year = 2024,
        month = oct,
       volume = {534},
       number = {1},
        pages = {325-348},
          doi = {10.1093/mnras/stae2092},
archivePrefix = {arXiv},
       eprint = {2405.02242},
 primaryClass = {astro-ph.GA},
       adsurl = {https://ui.adsabs.harvard.edu/abs/2024MNRAS.534..325C}
}

@ARTICLE{Aird2022,
       author = {{Aird}, James and {Coil}, Alison L. and {Kocevski}, Dale D.},
        title = "{AGN accretion and black hole growth across compact and extended galaxy evolution phases}",
      journal = {\mnras},
         year = 2022,
        month = oct,
       volume = {515},
       number = {4},
        pages = {4860-4889},
          doi = {10.1093/mnras/stac2103},
archivePrefix = {arXiv},
       eprint = {2201.11756},
 primaryClass = {astro-ph.GA},
       adsurl = {https://ui.adsabs.harvard.edu/abs/2022MNRAS.515.4860A}
}

@ARTICLE{Weller2024,
       author = {{Weller}, E. J. and {Pacucci}, F. and {Ni}, Y. and {Hernquist}, L. and {Park}, M.},
        title = "{Discrepancies between JWST Observations and Simulations of Quenched Massive Galaxies at z {\ensuremath{>}} 3: A Comparative Study with IllustrisTNG and ASTRID}",
      journal = {\apj},
         year = 2025,
        month = feb,
       volume = {979},
       number = {2},
        pages = {181},
          doi = {10.3847/1538-4357/ada360},
archivePrefix = {arXiv},
       eprint = {2406.02664},
 primaryClass = {astro-ph.GA}
}

@ARTICLE{Ni2024,
       author = {{Ni}, Y. and {Chen}, N. and {Zhou}, Y. and {Park}, M. and {Yang}, Y. and {Di Matteo}, T. and {Bird}, S. and {Croft}, R.},
        title = "{The Astrid Simulation: Evolution of Black Holes and Galaxies to z = 0.5 and Different Evolution Pathways for Galaxy Quenching}",
      journal = {\apj},
         year = 2025,
        month = sep,
       volume = {990},
       number = {2},
        pages = {120},
          doi = {10.3847/1538-4357/adf3a7},
archivePrefix = {arXiv},
       eprint = {2409.10666},
 primaryClass = {astro-ph.GA}
}

@ARTICLE{deGraaff2024,
       author = {{de Graaff}, A. and {Brammer}, G. and {Weibel}, A. and {Lewis}, Z. and {Maseda}, M. V. and {Oesch}, P. A. and {Bezanson}, R. and {Boogaard}, L. A. and {Cleri}, N. J. and {Cooper}, O. R. and et al.},
        title = "{RUBIES: A complete census of the bright and red distant Universe with JWST/NIRSpec}",
      journal = {\aap},
         year = 2025,
        month = may,
       volume = {697},
        pages = {A189},
          doi = {10.1051/0004-6361/202452186},
archivePrefix = {arXiv},
       eprint = {2409.05948},
 primaryClass = {astro-ph.GA}
}

@ARTICLE{Greene2005,
       author = {{Greene}, Jenny E. and {Ho}, Luis C.},
        title = "{Estimating Black Hole Masses in Active Galaxies Using the H{\ensuremath{\alpha}} Emission Line}",
      journal = {\apj},
         year = 2005,
        month = sep,
       volume = {630},
       number = {1},
        pages = {122-129},
          doi = {10.1086/431897},
archivePrefix = {arXiv},
       eprint = {astro-ph/0508335},
 primaryClass = {astro-ph},
       adsurl = {https://ui.adsabs.harvard.edu/abs/2005ApJ...630..122G}
}

@ARTICLE{Greene2024,
       author = {{Greene}, Jenny E. and {Labbe}, Ivo and {Goulding}, Andy D. and {Furtak}, Lukas J. and {Chemerynska}, Iryna and {Kokorev}, Vasily and {Dayal}, Pratika and {Volonteri}, Marta and {Williams}, Christina C. and {Wang}, Bingjie and {Setton}, David J. and {Burgasser}, Adam J. and {Bezanson}, Rachel and {Atek}, Hakim and {Brammer}, Gabriel and {Cutler}, Sam E. and {Feldmann}, Robert and {Fujimoto}, Seiji and {Glazebrook}, Karl and {de Graaff}, Anna and {Khullar}, Gourav and {Leja}, Joel and {Marchesini}, Danilo and {Maseda}, Michael V. and {Matthee}, Jorryt and {Miller}, Tim B. and {Naidu}, Rohan P. and {Nanayakkara}, Themiya and {Oesch}, Pascal A. and {Pan}, Richard and {Papovich}, Casey and {Price}, Sedona H. and {van Dokkum}, Pieter and {Weaver}, John R. and {Whitaker}, Katherine E. and {Zitrin}, Adi},
        title = "{UNCOVER Spectroscopy Confirms the Surprising Ubiquity of Active Galactic Nuclei in Red Sources at z > 5}",
      journal = {\apj},
         year = 2024,
        month = mar,
       volume = {964},
       number = {1},
          eid = {39},
        pages = {39},
          doi = {10.3847/1538-4357/ad1e5f},
archivePrefix = {arXiv},
       eprint = {2309.05714},
 primaryClass = {astro-ph.GA},
       adsurl = {https://ui.adsabs.harvard.edu/abs/2024ApJ...964...39G}
}

@ARTICLE{cigale,
       author = {{Yang}, G. and {Boquien}, M. and {Buat}, V. and {Burgarella}, D. and {Ciesla}, L. and {Duras}, F. and {Stalevski}, M. and {Brandt}, W.~N. and {Papovich}, C.},
        title = "{X-CIGALE: Fitting AGN/galaxy SEDs from X-ray to infrared}",
      journal = {\mnras},
         year = 2020,
        month = jan,
       volume = {491},
       number = {1},
        pages = {740-757},
          doi = {10.1093/mnras/stz3001},
archivePrefix = {arXiv},
       eprint = {2001.08263},
 primaryClass = {astro-ph.GA},
       adsurl = {https://ui.adsabs.harvard.edu/abs/2020MNRAS.491..740Y}
}

@ARTICLE{Harikane2023,
       author = {{Harikane}, Yuichi and {Zhang}, Yechi and {Nakajima}, Kimihiko and {Ouchi}, Masami and {Isobe}, Yuki and {Ono}, Yoshiaki and {Hatano}, Shun and {Xu}, Yi and {Umeda}, Hiroya},
        title = "{A JWST/NIRSpec First Census of Broad-line AGNs at z = 4-7: Detection of 10 Faint AGNs with M $_{BH}$ {}10$^{6}$-{}10$^{8}$ M $_{{\ensuremath{\odot}}}$ and Their Host Galaxy Properties}",
      journal = {\apj},
         year = 2023,
        month = dec,
       volume = {959},
       number = {1},
          eid = {39},
        pages = {39},
          doi = {10.3847/1538-4357/ad029e},
archivePrefix = {arXiv},
       eprint = {2303.11946},
 primaryClass = {astro-ph.GA},
       adsurl = {https://ui.adsabs.harvard.edu/abs/2023ApJ...959...39H}
}

@ARTICLE{Maiolino2023,
       author = {{Maiolino}, R. and {Scholtz}, J. and {Curtis-Lake}, E. and {Carniani}, S. and {Baker}, W. and {de Graaff}, A. and {Tacchella}, S. and {{\"U}bler}, H. and {D'Eugenio}, F. and {Witstok}, J. and et al.},
        title = "{JADES. The diverse population of infant black holes at 4 {\ensuremath{<}} z {\ensuremath{<}} 11: Merging, tiny, poor, but mighty}",
      journal = {\aap},
         year = 2024,
        month = nov,
       volume = {691},
        pages = {A145},
          doi = {10.1051/0004-6361/202347640},
archivePrefix = {arXiv},
       eprint = {2308.01230},
 primaryClass = {astro-ph.GA}
}

@ARTICLE{Kormendy2013,
       author = {{Kormendy}, John and {Ho}, Luis C.},
        title = "{Coevolution (Or Not) of Supermassive Black Holes and Host Galaxies}",
      journal = {\araa},
         year = 2013,
        month = aug,
       volume = {51},
       number = {1},
        pages = {511-653},
          doi = {10.1146/annurev-astro-082708-101811},
archivePrefix = {arXiv},
       eprint = {1304.7762},
 primaryClass = {astro-ph.CO},
       adsurl = {https://ui.adsabs.harvard.edu/abs/2013ARA&A..51..511K}
}

@ARTICLE{Carnall2023nat,
       author = {{Carnall}, Adam C. and {McLure}, Ross J. and {Dunlop}, James S. and {McLeod}, Derek J. and {Wild}, Vivienne and {Cullen}, Fergus and {Magee}, Dan and {Begley}, Ryan and {Cimatti}, Andrea and {Donnan}, Callum T. and {Hamadouche}, Massissilia L. and {Jewell}, Sophie M. and {Walker}, Sam},
        title = "{A massive quiescent galaxy at redshift 4.658}",
      journal = {\nat},
         year = 2023,
        month = jul,
       volume = {619},
       number = {7971},
        pages = {716-719},
          doi = {10.1038/s41586-023-06158-6},
archivePrefix = {arXiv},
       eprint = {2301.11413},
 primaryClass = {astro-ph.GA},
       adsurl = {https://ui.adsabs.harvard.edu/abs/2023Natur.619..716C}
}

@ARTICLE{Habouzit2021,
       author = {{Habouzit}, M{\'e}lanie and {Li}, Yuan and {Somerville}, Rachel S. and {Genel}, Shy and {Pillepich}, Annalisa and {Volonteri}, Marta and {Dav{\'e}}, Romeel and {Rosas-Guevara}, Yetli and {McAlpine}, Stuart and {Peirani}, S{\'e}bastien and {Hernquist}, Lars and {Angl{\'e}s-Alc{\'a}zar}, Daniel and {Reines}, Amy and {Bower}, Richard and {Dubois}, Yohan and {Nelson}, Dylan and {Pichon}, Christophe and {Vogelsberger}, Mark},
        title = "{Supermassive black holes in cosmological simulations I: M$_{BH}$ - M$_{{\ensuremath{\star}}}$ relation and black hole mass function}",
      journal = {\mnras},
         year = 2021,
        month = may,
       volume = {503},
       number = {2},
        pages = {1940-1975},
          doi = {10.1093/mnras/stab496},
archivePrefix = {arXiv},
       eprint = {2006.10094},
 primaryClass = {astro-ph.GA},
       adsurl = {https://ui.adsabs.harvard.edu/abs/2021MNRAS.503.1940H}
}

@ARTICLE{Stone2024,
       author = {{Stone}, Meredith A. and {Lyu}, Jianwei and {Rieke}, George H. and {Alberts}, Stacey and {Hainline}, Kevin N.},
        title = "{Undermassive Host Galaxies of Five z {\ensuremath{\sim}} 6 Luminous Quasars Detected with JWST}",
      journal = {\apj},
         year = 2024,
        month = mar,
       volume = {964},
       number = {1},
          eid = {90},
        pages = {90},
          doi = {10.3847/1538-4357/ad2a57},
archivePrefix = {arXiv},
       eprint = {2310.18395},
 primaryClass = {astro-ph.GA},
       adsurl = {https://ui.adsabs.harvard.edu/abs/2024ApJ...964...90S}
}

@ARTICLE{Greene2020,
       author = {{Greene}, Jenny E. and {Strader}, Jay and {Ho}, Luis C.},
        title = "{Intermediate-Mass Black Holes}",
      journal = {\araa},
         year = 2020,
        month = aug,
       volume = {58},
        pages = {257-312},
          doi = {10.1146/annurev-astro-032620-021835},
archivePrefix = {arXiv},
       eprint = {1911.09678},
 primaryClass = {astro-ph.GA},
       adsurl = {https://ui.adsabs.harvard.edu/abs/2020ARA&A..58..257G}
}

@ARTICLE{Kocevski2023,
       author = {{Kocevski}, D. D. and {Onoue}, M. and {Inayoshi}, K. and {Trump}, J. R. and {Arrabal Haro}, P. and {Grazian}, A. and {Dickinson}, M. and {Finkelstein}, S. L. and {Kartaltepe}, J. S. and {Hirschmann}, M. and et al.},
        title = "{Hidden Little Monsters: Spectroscopic Identification of Low-mass, Broad-line AGNs at z {\ensuremath{>}} 5 with CEERS}",
      journal = {\apjl},
         year = 2023,
        month = sep,
       volume = {954},
       number = {1},
        pages = {L4},
          doi = {10.3847/2041-8213/ace5a0},
archivePrefix = {arXiv},
       eprint = {2302.00012},
 primaryClass = {astro-ph.GA}
}

@ARTICLE{Pacucci2023,
       author = {{Pacucci}, F. and {Nguyen}, B. and {Carniani}, S. and {Maiolino}, R. and {Fan}, X.},
        title = "{JWST CEERS and JADES Active Galaxies at z = 4--7 Violate the Local M$_{\bullet}$--M$_{\star}$ Relation at {\ensuremath{>}}3{\ensuremath{\sigma}}: Implications for Low-mass Black Holes and Seeding Models}",
      journal = {\apjl},
         year = 2023,
        month = nov,
       volume = {957},
       number = {1},
        pages = {L3},
          doi = {10.3847/2041-8213/ad0158},
archivePrefix = {arXiv},
       eprint = {2308.12331},
 primaryClass = {astro-ph.GA}
}

@ARTICLE{Ubler2023,
       author = {{{\"U}bler}, H. and {Maiolino}, R. and {Curtis-Lake}, E. and {P{\'e}rez-Gonz{\'a}lez}, P. G. and {Curti}, M. and {Perna}, M. and {Arribas}, S. and {Charlot}, S. and {Marshall}, M. A. and {D'Eugenio}, F. and et al.},
        title = "{GA-NIFS: A massive black hole in a low-metallicity AGN at z {\ensuremath{\sim}} 5.55 revealed by JWST/NIRSpec IFS}",
      journal = {\aap},
         year = 2023,
        month = sep,
       volume = {677},
        pages = {A145},
          doi = {10.1051/0004-6361/202346137},
archivePrefix = {arXiv},
       eprint = {2302.06647},
 primaryClass = {astro-ph.GA}
}

@ARTICLE{Juodzbalis2026census,
       author = {{Juod{\v{z}}balis}, I. and {Maiolino}, R. and {Baker}, W. M. and {Lake}, E. C. and {Scholtz}, J. and {D'Eugenio}, F. and {Trefoloni}, B. and {Isobe}, Y. and {Tacchella}, S. and {Bunker}, A. J. and et al.},
        title = "{JADES: comprehensive census of broad-line AGN from reionization to cosmic noon revealed by JWST}",
      journal = {\mnras},
         year = 2026,
        month = feb,
       volume = {546},
       number = {3},
          eid = {stag086},
        pages = {stag086},
          doi = {10.1093/mnras/stag086},
archivePrefix = {arXiv},
       eprint = {2504.03551},
 primaryClass = {astro-ph.GA}
}

@ARTICLE{Kokorev2023,
       author = {{Kokorev}, V. and {Fujimoto}, S. and {Labbe}, I. and {Greene}, J. E. and {Bezanson}, R. and {Dayal}, P. and {Nelson}, E. J. and {Atek}, H. and {Brammer}, G. and {Caputi}, K. I. and et al.},
        title = "{UNCOVER: A NIRSpec Identification of a Broad-line AGN at z = 8.50}",
      journal = {\apjl},
         year = 2023,
        month = nov,
       volume = {957},
       number = {1},
        pages = {L7},
          doi = {10.3847/2041-8213/ad037a},
archivePrefix = {arXiv},
       eprint = {2308.11610},
 primaryClass = {astro-ph.GA}
}

@ARTICLE{Ren2025,
       author = {{Ren}, W. and {Silverman}, J. D. and {Faisst}, A. L. and {Fujimoto}, S. and {Yan}, L. and {Liu}, Z. and {Tsujita}, A. and {Aravena}, M. and {Davies}, R. L. and {De Looze}, I. and et al.},
        title = "{The ALPINE--CRISTAL-- JWST survey: revealing less massive black holes in high-redshift galaxies}",
      journal = {\mnras},
         year = 2025,
        month = oct,
       volume = {544},
       number = {1},
        pages = {211--233},
          doi = {10.1093/mnras/staf1709},
archivePrefix = {arXiv},
       eprint = {2509.02027},
 primaryClass = {astro-ph.GA}
}

@ARTICLE{Geris2026,
       author = {{Geris}, S. and {Maiolino}, R. and {Isobe}, Y. and {Scholtz}, J. and {D'Eugenio}, F. and {Ji}, X. and {Juod{\v{z}}balis}, I. and {Simmonds}, C. and {Dayal}, P. and {Trinca}, A. and et al.},
        title = "{JADES reveals a large population of low-mass black holes at high redshift}",
      journal = {\mnras},
         year = 2026,
       volume = {545},
       number = {1},
          eid = {staf1979},
        pages = {staf1979},
          doi = {10.1093/mnras/staf1979},
archivePrefix = {arXiv},
       eprint = {2506.22147},
 primaryClass = {astro-ph.GA}
}

@ARTICLE{Ziparo2026,
       author = {{Ziparo}, F. and {Carniani}, S. and {Gallerani}, S. and {Trefoloni}, B.},
        title = "{A selection-aware view of black hole--galaxy coevolution at high redshift}",
      journal = {\aap},
         year = 2026,
        month = oct,
       volume = {714},
        pages = {A290},
          doi = {10.1051/0004-6361/202659574},
archivePrefix = {arXiv},
       eprint = {2603.04358},
 primaryClass = {astro-ph.GA}
}

@ARTICLE{Brooks2026,
       author = {{Brooks}, M. and {Trump}, J. R. and {Simons}, R. C. and {Cole}, J. and {Taylor}, A. J. and {Bagley}, M. B. and {Finkelstein}, S. L. and {Davis}, K. and {Amor{\'{\i}}n}, R. O. and {Backhaus}, B. E. and et al.},
        title = "{Beyond the Monsters: A More Complete Census of Black Hole Activity at Cosmic Dawn}",
      journal = {\apj},
         year = 2026,
        month = may,
       volume = {1002},
       number = {2},
        pages = {129},
          doi = {10.3847/1538-4357/ae5652},
archivePrefix = {arXiv},
       eprint = {2511.19609},
 primaryClass = {astro-ph.GA}
}

@ARTICLE{SunY2025,
       author = {{Sun}, Y. and {Rieke}, G. H. and {Lyu}, J. . and {Stone}, M. A. and {Ji}, Z. and {Rinaldi}, P. and {Willmer}, C. N. A. and {Zhu}, Y.},
        title = "{Evolution of the M$_{*}$-- M$_{\rm BH}$ Relation from z {\ensuremath{\sim}} 6 to the Present Epoch}",
      journal = {\apj},
         year = 2025,
        month = apr,
       volume = {983},
       number = {2},
        pages = {165},
          doi = {10.3847/1538-4357/adc250},
archivePrefix = {arXiv},
       eprint = {2503.03675},
 primaryClass = {astro-ph.GA}
}

@ARTICLE{Zhuang2026,
       author = {{Zhuang}, M.-Y. and {Li}, J. and {Shen}, Y. and {Lin}, X. and {Shapley}, A. E. and {Wang}, F. and {Wu}, Q. and {Yang}, Q.},
        title = "{NEXUS: A Spectroscopic Census of Broad-line AGNs and Little Red Dots at 3 {\ensuremath{\lesssim}} z {\ensuremath{\lesssim}} 6}",
      journal = {\apj},
         year = 2026,
        month = mar,
       volume = {999},
       number = {1},
        pages = {31},
          doi = {10.3847/1538-4357/ae3612},
archivePrefix = {arXiv},
       eprint = {2505.20393},
 primaryClass = {astro-ph.GA}
}

@ARTICLE{Stone2025,
       author = {{Stone}, M. A. and {Rieke}, G. H. and {Lyu}, J. . and {Florian}, M. K. and {Hainline}, K. N. and {Sun}, Y. and {Zhu}, Y.},
        title = "{The z = 7.08 Quasar ULAS J1120+0641 May Never Reach a ``Normal'' Black Hole to Stellar Mass Ratio}",
      journal = {\apj},
         year = 2025,
        month = nov,
       volume = {993},
       number = {2},
        pages = {168},
          doi = {10.3847/1538-4357/ae062a},
archivePrefix = {arXiv},
       eprint = {2507.13489},
 primaryClass = {astro-ph.GA}
}

@ARTICLE{Koudmani2026,
       author = {{Koudmani}, S. and {Scholtz}, J. and {Taylor}, A. J. and {Juod{\v{z}}balis}, I. and {Sijacki}, D. and {Somerville}, R. S. and {Maiolino}, R. and {Curtis-Lake}, E. and {Finkelstein}, S. L. and {Bourne}, M. A. and et al.},
        title = "{How to raise a supermassive black hole: interpreting early JWST AGN with the AESOPICA simulations}",
      journal = {arXiv e-prints},
         year = 2026,
        month = jul,
          eid = {arXiv:2607.26177},
        pages = {arXiv:2607.26177},
          doi = {10.48550/arXiv.2607.26177},
archivePrefix = {arXiv},
       eprint = {2607.26177},
 primaryClass = {astro-ph.GA}
}

@ARTICLE{Li2022,
       author = {{Li}, J. and {Silverman}, J. D. and {Izumi}, T. and {He}, W. and {Akiyama}, M. and {Inayoshi}, K. and {Matsuoka}, Y. and {Onoue}, M. and {Toba}, Y.},
        title = "{On the Connection between Supermassive Black Holes and Galaxy Growth in the Reionization Epoch}",
      journal = {\apjl},
         year = 2022,
        month = may,
       volume = {931},
       number = {1},
        pages = {L11},
          doi = {10.3847/2041-8213/ac6de8},
archivePrefix = {arXiv},
       eprint = {2203.10663},
 primaryClass = {astro-ph.GA}
}

@ARTICLE{Habouzit2022b,
       author = {{Habouzit}, M. and {Onoue}, M. and {Ba{\~n}ados}, E. and {Neeleman}, M. and {Angl{\'e}s-Alc{\'a}zar}, D. and {Walter}, F. and {Pillepich}, A. and {Dav{\'e}}, R. and {Jahnke}, K. and {Dubois}, Y.},
        title = "{Co-evolution of massive black holes and their host galaxies at high redshift: discrepancies from six cosmological simulations and the key role of JWST}",
      journal = {\mnras},
         year = 2022,
        month = feb,
       volume = {511},
       number = {3},
        pages = {3751--3767},
          doi = {10.1093/mnras/stac225},
archivePrefix = {arXiv},
       eprint = {2201.09892},
 primaryClass = {astro-ph.CO}
}

\end{document}